\documentclass[a4paper, 8pt]{article}
\usepackage[a4paper, left=1in, right=1in, top=1in, bottom=1in, footskip=.25in]{geometry}
\usepackage[dvipsnames]{xcolor}
\usepackage[utf8]{inputenc}
\usepackage{amsmath, amssymb}
\usepackage{subcaption}
\allowdisplaybreaks[4]
\usepackage{graphicx}
\usepackage{hyperref}
\usepackage{float}
\usepackage{multirow}
\usepackage{multicol}
\usepackage{booktabs}
\usepackage{makecell}
\usepackage{amsmath, mathtools, nccmath}
\title{Robust Parameter Estimation in Dynamical Systems by Stochastic Differential Equations}
\author{Qingchuan Sun$^{1,2}$, Susanne Ditlevsen$^{2}$\\1. Bielefeld Graduate School of Economics and Management, \\University of Bielefeld\\ 2. Department of Mathematical Sciences, University of Copenhagen,\\ Copenhagen, Denmark\\
Emails: QS: qingchuan.sun@uni-bielefeld.de, SD: susanne@math.ku.dk}

\date{}

\begin{document}

\maketitle

\begin{abstract}
Ordinary and stochastic differential equations (ODEs and SDEs) are widely used to model continuous-time processes across various scientific fields. While ODEs offer interpretability and simplicity, SDEs incorporate randomness, providing robustness to noise and model misspecifications. Recent research highlights the statistical advantages of SDEs, such as improved parameter identifiability and stability under perturbations. This paper investigates the robustness of parameter estimation in SDEs versus ODEs under three types of model misspecifications: unrecognized noise sources, external perturbations, and simplified models. Furthermore, the effect of missing data is explored. Through simulations and an analysis of Danish COVID-19 data, we demonstrate that SDEs yield more stable and reliable parameter estimates, making them a strong alternative to traditional ODE modeling in the presence of uncertainty.
\end{abstract}

\section{Introduction}
Dynamical systems described by ordinary and stochastic differential equations (ODEs and SDEs) are popular models for continuous-time processes, with applications in many fields. Examples of classic ODE models are the Lotka–Volterra equations in population dynamics, master equations in chemistry, compartment models in epidemiology, and ODEs are used to model many complex systems due to their concise form and physical interpretation. Also SDEs have many applications, for example in finance, biology, physics, epidemiology and climate science, where system noise is included in the model, thus affecting the future evolution of the process.


Recent studies have shown that SDE models have statistical advantages over ODE models. In kinetic modeling of translation kinetics after mRNA transfection, the SDE model provides better parameter identifiability than the ODE model \cite{SPieschner}. In Neural SDEs, the inclusion of noise terms improves the robustness of the model to perturbations in state predictions compared to Neural ODEs \cite{Liu:2020}. Loss functions in nonlinear ODEs often suffer from multiple optimas jeopardizing parameter estimation, and SDEs provide a tool to regularize the objective function, thus reducing the number of local minima to obtain more stable parameter estimates \cite{Leander:2014}. In \cite{ditlevsen:2020}, it was shown that for the class of SDEs called Pearson diffusions, rate parameters can be estimated more accurately by external perturbations. 

In practice, the true dynamics of a process is always unknown. Measurement errors are a common source of randomness. However, other types of systematic errors and perturbations induced from outside also affect the dynamics and future evolution of the process, as well as deviations in the governing equations. These factors are often unknown, cannot be measured and are thus difficult or impossible to include in the statistical model. SDE models provide good alternatives to ODE models for parameter estimation and prediction that are robust to such deviations, as we will show. Furthermore, parameter estimates of ODE models can be sensitive to the number of observations in a dramatic way and single observations might drive the inference, while SDE models provide relatively stable estimates \cite{Cuenod:2011}, offering more robust statistical inference. For example, in \cite{Donnet:2010}, the SDE model gives more robust posterior predictive distributions compared to the ODE model in an analysis of chicken growth curves.

Parameter estimation has always been an important part of statistical inference. Data are often generated from a mix between known physical laws and a variety of complex and unknown disturbances. It is therefore critical to obtain robust parameter estimation procedures also under some sort of model misspecification. We aim for statistical models that can withstand perturbations and deviations from the ideal model and still provide valuable information about the dynamics.

Many parameter estimation methods have been developed for ODE and SDE models. Least Squares Estimation (LSE) is probably the most commonly used in ODE models, and coincides with the maximum likelihood estimator (MLE) if the measurement noise is normally distributed. A significant amount of research has been devoted to improving LSE by developing fast convergence methods to derive parameters efficiently. These methods often involve iterative algorithms, such as gradient-based optimization techniques, including gradient descent and its variants, such as the Levenberg-Marquardt algorithm \cite{Jain2017, Transtrum2012, 10.5555/3157096.3157142}. Bayesian inference has also been used to estimate parameters, providing alternatives to account for uncertainty and prior knowledge of parameters \cite{GuedjJ2007MLEi, Huang2020}.

The likelihood function is often unknown for SDE models, and thus, the MLE is rarely feasible, only for linear models and a few other models is the likelihood available. Bayesian methods are an alternative to MLE, see \cite{fuchsbook, SDEbook2019} for comprehensive reviews of estimation methods for SDEs. Approximations to the MLE are also popular, constructing so-called pseudo-likelihoods. 
Many methods have been developed to construct pseudo-likelihood functions, such as the Euler-Maruyama (EM) approximation, the extended Kalman filter (EKF)\cite{OuLu2023Eonm}, the Unscented Kalman filter (UKF)\cite{li2022constrained}, and the Strang splitting scheme \cite{pilipovic:2021} for nonlinear models. We will use the latter in Section \ref{Example 2}. 

In this paper, we explore the quality and robustness of parameter estimates of SDE models compared to ODE models subject to different model misspecifications. The model misspecifications we consider can be grouped into three types: 1. Wrongly identifying the random sources, which are either measurement errors or system noise present in the dynamics; 2. External perturbations; and 3. Simplified models, where the data are generated from dynamics that are more complex than the fitted model.

In Section \ref{Models}, we introduce the ODE and SDE models and two examples, a simple one-dimensional linear model and two epidemic models, the SIR and SEIR models. We introduce different types of perturbation and model misspecifications to evaluate and compare the performance of the ODE and SDE models in response to external perturbations and data from more complex models in Section \ref{Model misspecification}. In Section \ref{Simulations}, we present a simulation study of the two examples to illustrate the performance of ODE and SDE models, and study the possible consequences of model misspecification. In Section \ref{application}, we analyse COVID-19 data from Denmark and check the stability of parameter estimates for data sets with missing data. We change the start and end times of the data set and investigate the quality of the parameter estimates given by the deterministic and stochastic SIR and SEIR models. Finally, we conclude in Section \ref{Conclusion} and provide some further technical results in Section \ref{Calculation}.

\section{Models}\label{Models}
Consider a $d$-dimensional continuous dynamical system with state vector $\mathbf{x}_t^{ode} \in \mathbb{R}^d$ described by an ODE,
\begin{align}
\label{ODE general}
    d \mathbf{x}_t^{ode} = \mathbf{f}(\mathbf{x}_t^{ode},\mathbf{\theta}) dt, \qquad \mathbf{x}_{t_0}^{ode} = \mathbf{x}_0,
\end{align}
where $t \geq t_0 \in \mathbb{R}$ denotes time. We wish to estimate the  parameter vector $\mathbf{\theta}\in \Theta \subseteq\mathbb{R}^p$ belonging to some $p$-dimensional parameter space from discrete observations corrupted by Gaussian measurement noise, 
\begin{align}
\label{ODE measurement}
   \mathbf{y}_k^{ode} = \mathbf{x}_k^{ode}  + \mathbf{e}_k, \quad \mathbf{e}_k \overset{\text{i.i.d}}{\sim} \mathcal{N}(0, \mathbf{\Sigma}_0)
\end{align}
measured at discrete time points $t_k, k = 0,1,\ldots , n$, where $\mathbf{\Sigma}_0$ is the covariance matrix of the measurement noise.

We extend the ODE model \eqref{ODE general} to an SDE model by adding system noise of the form
\begin{align}
\label{SDE general}
   d \mathbf{x}_t^{sde} = \mathbf{f}(\mathbf{x}_t^{sde},\theta) dt + \mathbf{\Sigma} d\mathbf{W}_t, \qquad \mathbf{x}_{t_0}^{sde} = \mathbf{x}_0,
\end{align}
where $\mathbf{\Sigma} \in \mathbb{R}^{d \times q}, q \leq d$ is the diffusion coefficient and $\mathbf{W}_t$ is a $q$-dimensional Brownian motion. The observations are assumed without measurement noise,
\begin{align}
\label{SDE measurement}
   \mathbf{y}_k^{sde} = \mathbf{x}_k^{sde}.
\end{align}
The goal is to investigate the effect of how the noise is modeled on the statistical properties of the estimators of $\theta$ under different data-generating models, possibly not agreeing with the statistical model. We therefore do not consider measurement error in the SDE model, such that deviations from the drift $\mathbf{f}$ is either caused by measurement error (ODE model) or system noise (SDE model). However, in Section \ref{application} we allow for measurement noise also in the SDE model when analysing COVID-19 data. The SDE model \eqref{SDE general} reduces to the ODE model \eqref{ODE general} for $\mathbf{\Sigma}=\mathbf{0}$. 

For simplicity, we assume equidistant time steps, i.e., $t_k = k\Delta, k= 0,\ldots,n$, where $\Delta = t_k - t_{k-1}$. The observation interval is thus $T = n\Delta$. We use superscripts $^{ode}$ and $^{sde}$ to indicate the model from which the data are sampled, as well as for the respective estimators assuming either one or the other model. We denote by $\mathbf{y}^{*} = ({y}^{*}_0, {y}^{*}_1, \ldots , {y}^{*}_n)$ the entire set of observations, where $*$ is $ode$ or $sde$. 

All simulated trajectories throughout the paper were simulated with an Euler-Maruyama (EM) scheme with step size $0.01$ and then subsampled to the relevant step size $\Delta$ to decrease discretization errors. 

\subsection{Example 1: Linear drift }\label{Example 1}
Consider $d=1$ and linear drift 
\begin{align}\label{drift example 1}
     \mathbf{f}(\mathbf{x},a,b) = -a(\mathbf{x}-b),
\end{align}
where $a>0$ denotes the rate at which the system moves towards the long-term level $b \in \mathbb{R}$. The parameters to estimate are $\theta = (a,b)$ and a variance parameter.

The ODE model \eqref{ODE general} has the form
\begin{eqnarray}\label{example 1 ODE}
    dx_t^{ode} &=& -a(x_t^{ode} - b)dt, \nonumber \\
    y_k^{ode} &=& x_k^{ode} + e_k, \quad e_k \overset{\text{i.i.d}}{\sim} \mathcal{N}(0, \sigma_0^2)
\end{eqnarray} 
for $k=0,\ldots , n$, so $\mathbf{\Sigma}_0 = \sigma_0^2$ in \eqref{ODE measurement}. 

The SDE model \eqref{SDE general} with drift \eqref{drift example 1} and diffusion coefficient $\mathbf{\Sigma}=\sigma$ is 
\begin{align}\label{example 1 SDE}
    dx_t^{sde} = -a(x_t^{sde} - b)dt + \sigma dW_t,
\end{align} 
which is an Ornstein-Uhlenbeck (OU) process and $y_k^{sde} = x_k^{sde}$. 

\begin{figure}[!t]
    \centering
    \includegraphics[width = 0.8\textwidth]{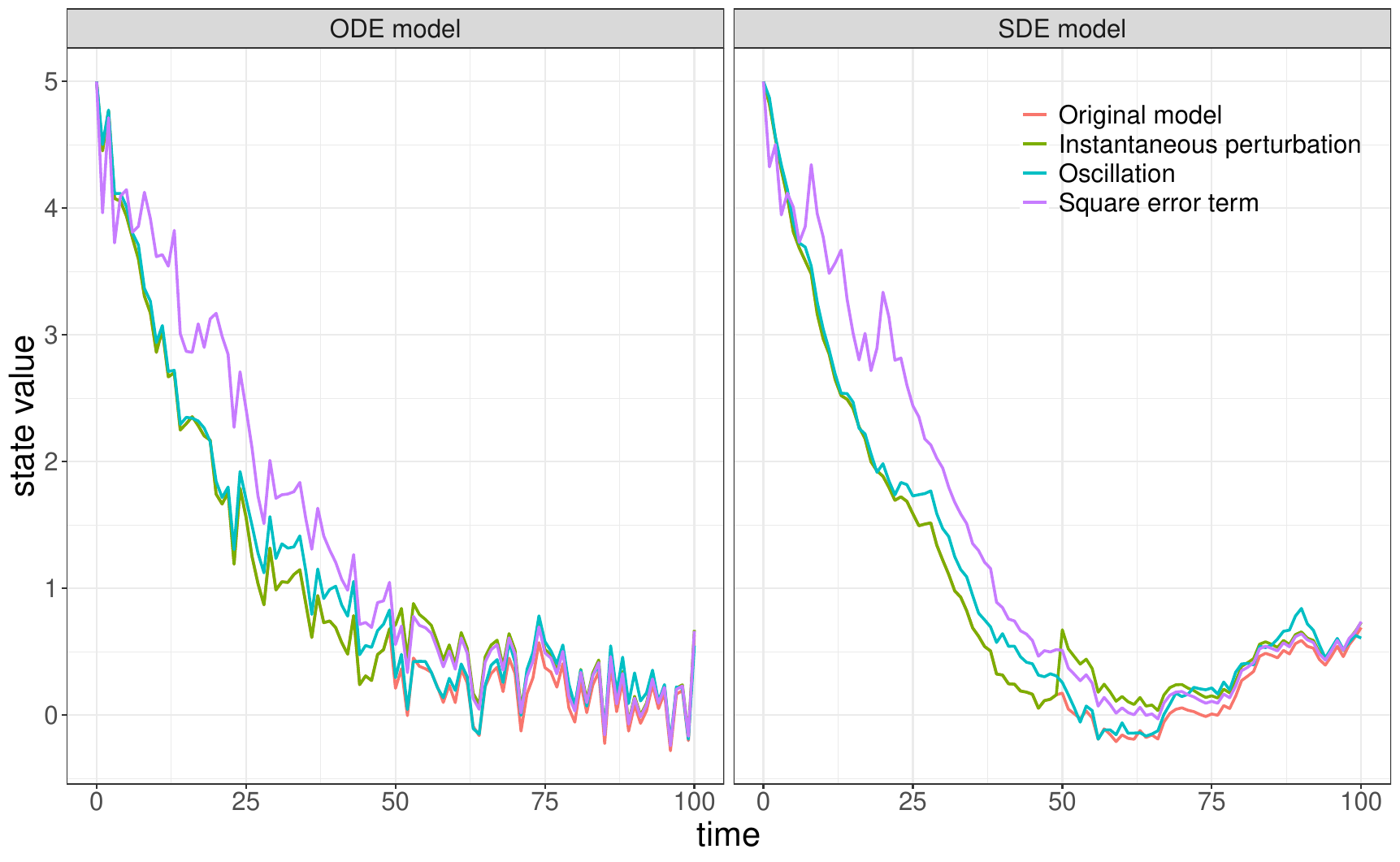} 
    \caption{{\bf Example trajectories of the linear model with perturbations.} Shown are the ODE model \eqref{example 1 ODE}, the SDE model \eqref{example 1 SDE} and models with three types of perturbation (see Section \ref{Estimators Example 1}) of time length $T = 100$ and time step $\Delta = 1$. We use the same set of random numbers to simulate the original and the perturbed models so the difference between trajectories are solely caused by the perturbation. The instantaneous perturbation is of size $h=0.5$ and happens at time $t_p = 50$. The unperturbed and the instantaneously perturbed trajectories are therefore the same up to time $t_p$. The parameters in the other two perturbation scenarios are $\sigma_b = 1$ and $\sigma_{\gamma} = 0.2$, respectively. Other parameters are $a=0.05$, $x_0 = 5$, $b=0$, $\sigma = 0.05$ and $\sigma_0 = \sigma/\sqrt{2a}$.}
    \label{fig: example 1 trajectory comparison}
\end{figure}

The conditional variance of $y_{k}^{sde}$ given $y_{k-1}^{sde}$ is $\text{V}(y_k^{sde}|y_{k-1}^{sde})=\sigma^2(1-\rho^{2})/{2a}$, with one-lag autocorrelation $\rho = e^{-a\Delta}$ and asymptotic variance $\sigma^2/2a$, see \cite{ditlevsen:2020}. We force the variances of $y^{ode}_k$ and $y^{sde}_k$ to be comparable by setting $\sigma_0^2 = \sigma^2/2a$ in the simulations. In Figure \ref{fig: example 1 trajectory comparison}, example trajectories of models \eqref{example 1 ODE} and \eqref{example 1 SDE} are illustrated. We set $n=100$ and $\Delta=1$, so that $T=100$. Parameters are $a=0.05, b=0, \sigma = 0.05$ and $\sigma_0 = \sigma/\sqrt{2a}$. 

\subsection{Example 2: The SIR epidemic model}\label{Example 2}
The SIR model \cite{Bartlett:1949, kermack:1927} is a compartmental model widely used in epidemiology, with three compartments representing different disease states: S (Susceptible), I (Infectious) and R (Removed or Recovered). It describes a possible mechanism for the spread of an infectious disease in a population. The compartment R includes those who recover from the disease or die. It is assumed that those who recover from the disease will not get infected again. We denote the population size of each compartment at time $t$ by $(S_t, I_t, R_t)$. The total population size $S_t+I_t+R_t$ is fixed to $N$. 

Instead of using the absolute population size in each compartment, we use proportions $(s_t,i_t,r_t) = (\frac{S_t}{N},\frac{I_t}{N},\frac{R_t}{N})$. The SIR model is given by
\begin{align}\label{model SIR}
    \begin{aligned}
        ds_t &= -\alpha s_t i_t dt + \sigma_1 dW_t^1,  \\
        di_t &= (\alpha s_t i_t - \beta i_t)dt + \sigma_2 dW_t^2,
    \end{aligned}
\end{align}
with $r_t = 1-s_t-i_t$, where $\alpha >0$ represents the contact rate between an infectious and a susceptible individual and $\beta >0$ is the removal rate from the infectious state. In the ODE model, $\sigma_1= \sigma_2 =0$, and in the SDE model, $\sigma_1, \sigma_2 \geq 0, \sigma_1+\sigma_2>0$ and $(W_t^1, W_t^2)$ are two independent standard Brownian motions. 

The measurements of the ODE model are
\begin{align}
\label{SIR deterministic error}
    \mathbf{y}_k^{ode} = 
    \begin{pmatrix}
        s_k\\
        i_k
    \end{pmatrix}
    +  \mathbf{e}_k , \qquad  \mathbf{e}_k \sim \mathcal{N} 
    \begin{pmatrix}
        \begin{pmatrix}
            0 \\
            0
        \end{pmatrix},
        \begin{pmatrix}
            \gamma_1^2 & 0 \\
            0 & \gamma_2^2
        \end{pmatrix}
    \end{pmatrix},
\end{align}
with $\gamma_1, \gamma_2 >0$. The measurements of the SDE model are
\begin{align}
\label{SIR stochastic error}
    \mathbf{y}_k^{sde} = 
    \begin{pmatrix}
        s_k\\
        i_k
    \end{pmatrix}.
\end{align}
The parameters to estimate are $\theta = (\alpha, \beta)$ and variance parameters $\gamma_i^2$ for the ODE model or $\sigma_i^2$ for the SDE model, $i=1,2$. To keep the errors of the two models at a similar level, we set their variances to have the following relationship
$$\gamma_1^2 = \text{Var}(s_T); \quad  \gamma_2^2 = \text{Var}(i_T),$$
where $\text{Var}(s_T) \approx \sigma_1^2 T$ and $\text{Var}(i_T) \approx \sigma_2^2/2( \beta- \alpha s^*)$, assuming that the deterministic SIR model has reached $(s^*, 0)$ at time $T$. We obtain $s^*$ numerically by solving the equation
$\frac{\alpha}{\beta} (1- s^*) + \ln \frac{s^*}{s_0} = 0$ (\cite{Piovella:2020}).

\begin{figure}[!ht]
    \centering
    \includegraphics[width = 0.8\textwidth]{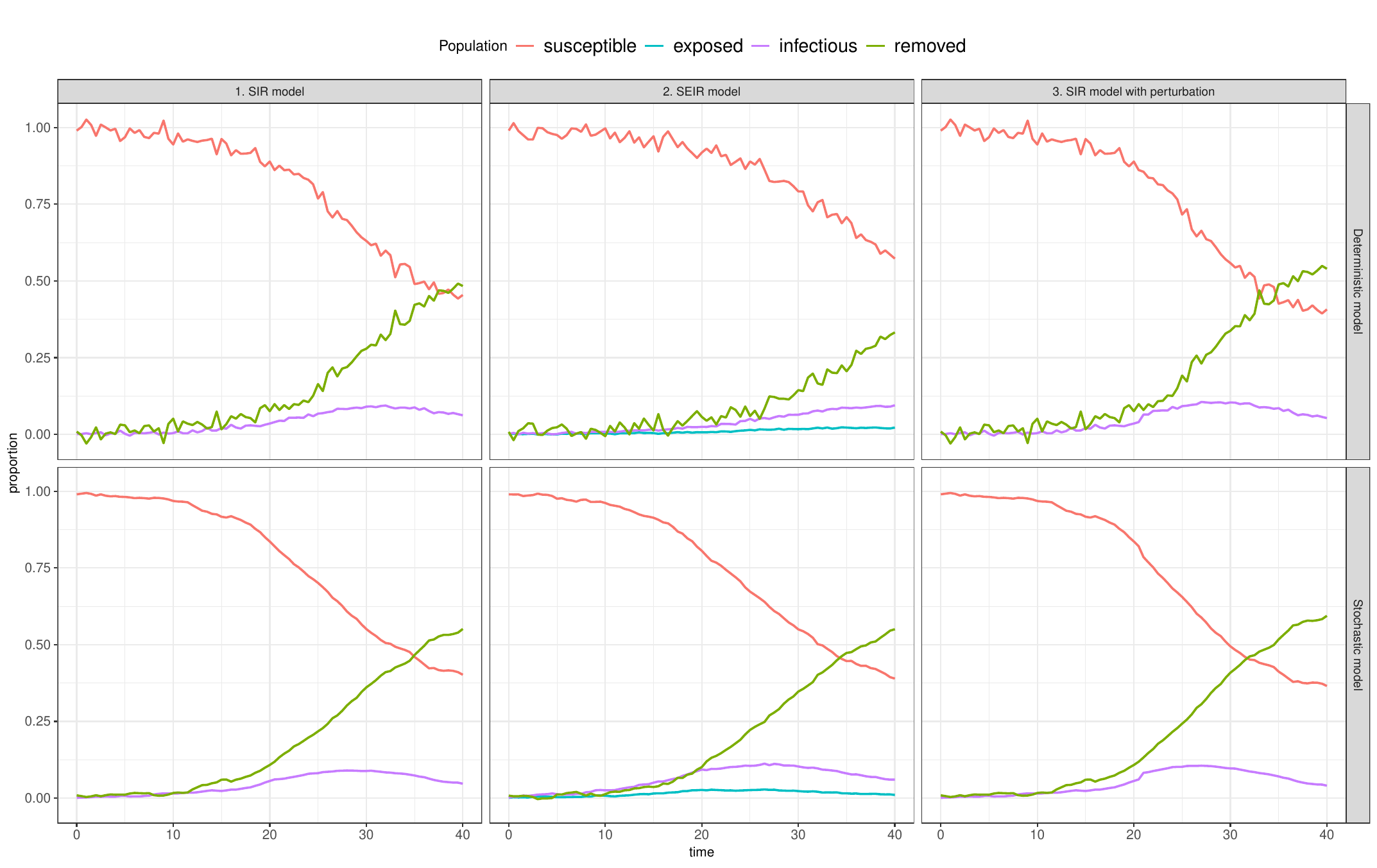} 
    \caption{{\bf Example trajectories of the epidemic models.} Shown are ODE and SDE versions of: 1. SIR model \eqref{model SIR}, 2. SEIR model \eqref{SEIR model} and 3. SIR model with instantaneous perturbation \eqref{model SIR} and \eqref{SIR perturbation}. Data were simulated with parameters $\alpha = 0.5$, $\lambda = 1$, $\beta = 0.3$. In the SIR model with and without instantaneous perturbation, we set $\sigma_1 = 5\times10^{-3}$, $\sigma_2 = 1\times10^{-3}$,   $\gamma_1 = \sigma_1\sqrt{T}$, $ \gamma_2 = \sigma_2/\sqrt{2(\beta - \alpha s^*)}$. In the SEIR model we set $\sigma_1 = 5\times10^{-3}$, $\sigma_2 = \sigma_3 = 1\times10^{-3}$,  $ \gamma_1 = \sigma_1\sqrt{T}$, $ \gamma_2 =\sigma_2/\sqrt{2\lambda}$, $ \gamma_3 = \sigma_3/\sqrt{2\beta}$. Initial values were $s_0 = 0.99$, $e_0 = i_0 = 0.001$.}
    \label{fig: SIR model}
\end{figure}

We show example trajectories of the deterministic and the stochastic SIR models in the first column of Figure \ref{fig: SIR model}. We set $n=80$, $T=40$ and $\Delta=0.5$. Parameters were set to $\alpha = 0.5$, $ \beta = 0.3$, $ \sigma_1 = 5\times10^{-3}$, $ \sigma_2 = 1\times10^{-3}$, $\gamma_1 = \sigma_1\sqrt{T}$, $ \gamma_2 = \sigma_2/\sqrt{2(\beta - \alpha s^*)}$ and initial values $s_0 = 0.99$, $ i_0 = 0.001$

\subsection{Estimators} \label{Estimators}
The impacts of the noise on the dynamics of the ODE and SDE models are different. In the ODE model, the dynamics evolves deterministically according to the solution of the ODE, and the measurement errors do not affect the future dynamics. In the SDE model, the randomness in the dynamics affects the subsequent states and makes the true process deviate from the ODE solution.

For the ODE model, we use non-linear LSE to estimate $\theta$, which corresponds to the MLE since the measurement errors are Gaussian. The drift parameters are estimated by minimising the sum of squares,
\begin{align}\label{lse formula}
    \hat{\theta} =  \mathop{\arg\min}_{\theta}  \sum_{k} \left(y_k^{ode}-\mathbf{F}(t_k,\theta)\right)^\top\left(y_k^{ode}-\mathbf{F}(t_k,\theta)\right),
\end{align}
where $\mathbf{F}(t, \theta)$ is the solution to the ODE \eqref{ODE general} (assuming it exists), and $^\top$ denotes transposition. The coordinates might need to be normalized such that the scale of one coordinate will not dominate the sum, however, this is not relevant in our examples. The variance parameters are then estimated from the residuals,
\begin{align}\label{lse sigma}
    \hat{\mathbf{\Sigma}}_0 =  \frac{1}{n-p}  \sum_{k} \left(y_k^{ode}-\mathbf{F}(t_k,\hat \theta)\right)\left(y_k^{ode}-\mathbf{F}(t_k,\hat \theta)\right)^\top.
\end{align}
There are no analytical forms for the LSEs for Examples 1 or 2. 

The ideal estimator for an SDE model is the MLE, however, the transition density is needed, which is unknown in most cases. It is available for the OU process in Example 1, but not for the SIR model in Example 2. 
The MLE of the OU process \eqref{example 1 SDE} is given as solutions to the following equations (see \cite{ditlevsen:2020} for details), 
\begin{align}
    \label{b_hat}
    \hat{b} &= \frac{1}{n}\sum_{k=1}^n y_k^{sde} + \frac{\hat{\rho}}{n(1-\hat{\rho})}(y_n^{sde} - y_0^{sde}), \\
     \label{a_hat}
    \hat{\rho} &= \frac{\sum_{k=1}^n (y_k^{sde}-\hat{b})(y_{k-1}^{sde}-\hat{b})}{\sum_{k=1}^n (y_{k-1}^{sde}-\hat{b})^2}, \\
    \label{sigma}
    \hat{\sigma}^{2} &= \frac{2\sum_{k=1}^n \hat{a} \left( (y_k^{sde} - y_{k-1}^{sde}\hat{\rho} -\hat{b}(1 -\hat{\rho}))\right)^2 }{n(1-\hat{\rho})},
\end{align}
where $\hat{a} = -\frac{1}{\Delta} \log\hat{\rho}$. 
    
For the stochastic SIR model, we use a pseudo-likelihood function as an approximation of the true likelihood. In \cite{pilipovic:2021}, a Strang splitting scheme was proposed to construct a pseudo-likelihood, which was shown to be consistent and asymptotically normal, outperforming other estimators on precision and computational speed. 

First, we partition the SDE into an OU process and a nonlinear ODE. The stochastic SIR model \eqref{model SIR} can be split as follows
\begin{align*}
    &\begin{aligned}
        d\mathbf{x}^{[1]}_t = \mathbf{A}\mathbf{x}^{[1]}_t dt + \mathbf{\Sigma}\mathbf{W}_t,
    \end{aligned}
    \\[1ex]
    &\begin{aligned}
        d\mathbf{x}^{[2]}_t = \mathbf{f}(\mathbf{x}^{[2]}_t)dt,
    \end{aligned}
\end{align*}
where $\mathbf{x}^{[1]}_t = (s^{[1]}_t, i^{[1]}_t)^\top$, $\mathbf{x}^{[2]}_t = (s^{[2]}_t, i^{[2]}_t)^\top$, $\mathbf{W}_t = (W^1_t, W^2_t)^\top$, and 
$$ \mathbf{A} =
   \begin{bmatrix}
    -\alpha & 0\\
    \alpha & -\beta
    \end{bmatrix},
    \quad \quad
    \mathbf{\Sigma} = 
     \begin{bmatrix}
    \sigma_1 & 0\\
    0 & \sigma_2
    \end{bmatrix},
    \quad \quad
    \mathbf{f}(\mathbf{x}^{[2]}_t) = 
    \begin{bmatrix}
    -\alpha s^{[2]}_t i^{[2]}_t + \alpha s^{[2]}_t \\
    \alpha s^{[2]}_t i^{[2]}_t - \alpha s^{[2]}_t
    \end{bmatrix}.
$$
We obtain the following two flows:
\begin{align*}
    &\begin{aligned}
       \mathbf{x}^{[1]}_{t+\Delta} = e^{\mathbf{A}\Delta} \mathbf{x}^{[1]}_{t} + \mathbf{\xi}_{\Delta},\\
    \end{aligned}
    \intertext{with $\mathbf{\xi}_{\Delta} \sim \mathcal{N}(0, \mathbf{\Omega}_{\Delta})$ where}    
    &\begin{aligned}\mathbf{\Omega}_\Delta &= \int_0^{\Delta} e^{\mathbf{A}(\Delta-u)}\mathbf{\Sigma}\mathbf{\Sigma}^\top e^{\mathbf{A}^\top(\Delta-u)} du \\
    &= 
           \sigma^2_1 \begin{bmatrix}
       \frac{1-e^{-2\alpha\Delta}}{2\alpha}& \frac{\alpha}{\beta-\alpha} \left ( \frac{1-e^{-2\alpha\Delta}}{2\alpha} - \frac{1-e^{-(\alpha+\beta)\Delta}}{\alpha + \beta} \right )\\
       \frac{\alpha}{\beta-\alpha} \left ( \frac{1-e^{-2\alpha\Delta}}{2\alpha} - \frac{1-e^{-(\alpha+\beta)\Delta}}{\alpha + \beta} \right ) &
        \frac{\alpha^2}{(\beta-\alpha)^2} \left ( \frac{1-e^{-2\alpha\Delta}}{2\alpha} - 2\frac{1-e^{-(\alpha+\beta)\Delta}}{\alpha + \beta} \right )+ \frac{\sigma^2_2}{\sigma^2_1} \frac{1-e^{-2\beta\Delta}}{2\beta}
       \end{bmatrix}
    \end{aligned}
    \intertext{and}
    &\begin{aligned}
       \mathbf{x}^{[2]}_{t+\Delta} =  \mathbf{F}(\mathbf{x}^{[2]}_{t}, \Delta)=
       \begin{bmatrix}
       \frac{c_1(t)-1}{1-e^{(\alpha \Delta + c_2(t))(c_1(t)-1)}}\\
       c_1(t) - \frac{c_1(t)-1}{1-e^{(\alpha \Delta + c_2(t))(c_1(t)-1)}}
       \end{bmatrix},
    \end{aligned}
\end{align*}
with 
$$ c_1(t)= s^{[2]}_{t}+ i^{[2]}_{t}, \qquad 
c_2(t) = \frac{1}{s^{[2]}_{t}+ i^{[2]}_{t} -1} \log \frac{1-i^{[2]}_{t}}{s^{[2]}_{t}}.$$
The flow of $\mathbf{x}^{[2]}_{t+\Delta}$ is defined for $c_1(t) < 1$, which is satisfied naturally in the SIR model. The pseudo-negative log-likelihood is
$$\ell(\alpha,\beta) = \frac{n}{2}\log(\det \mathbf{\Omega}_\Delta) + \frac{1}{2}\sum_{k=1}^n \mathbf{z}_{t_k}^T
\mathbf{\Omega}^{-1}_\Delta \mathbf{z}_{t_k}
- \sum_{k=1}^n \log \left|\det D \mathbf{F}^{-1} (\mathbf{x}_{t_k}, \Delta/2)  \right|,$$
with 
$$\mathbf{z}_{t_k} =  \mathbf{F}^{-1} (\mathbf{x}_{t_k}, \Delta/2) - e^{\mathbf{A}\Delta} \mathbf{F} (\mathbf{x}_{t_{k-1}}, \Delta/2),$$
for details, see \cite{pilipovic:2021}. Then the pseudo-MLE of the SIR model is computed as
\begin{equation}
    \label{Strang splitting scheme}
(\hat{\alpha},\hat{\beta}, \hat{\mathbf{\Sigma}}) :=  \mathop{\arg\min}_{\alpha,\beta, \mathbf{\Sigma}} \ell(\alpha,\beta, \mathbf{\Sigma}). 
\end{equation}

\section{Model misspecification}\label{Model misspecification}
We consider two forms of additive error, the measurement error in the ODE model and the internal system noise in the SDE model. However, it is difficult to determine whether the uncertainty in the data arises from the measurements or the dynamics themselves, or both. We first consider the impact of model misspecification of the noise on parameter estimation from finite samples. Then we introduce different types of perturbation to the trajectories. Instantaneous jumps in state and persistent state shifts are common patterns of perturbations seen in real world data; the former affects the dynamics by changing the state at a specific moment in time, while the latter not only affects the state but changes the parameters of the dynamics. The hypothesis is that fitting with an SDE model can compensate for the perturbations, whereas fitting with an ODE model will lead to biased estimates.

\subsection{Example 1: Finite samples and perturbations in the linear model}\label{Estimators Example 1}

\paragraph{Original model with no perturbation.}\label{Estimators Example 1 no perturbation}

We will now investigate possible biases in the parameter estimates from finite samples of size $n$, using the techniques proposed in \cite{ditlevsen2021, ditlevsen:2020}. First, we investigate the estimate of $\rho = e^{-a\Delta}$, which is the autocorrelation in the SDE model, estimated in \eqref{lse formula} if assuming an ODE model, and in \eqref{a_hat} assuming an SDE model. Given data generated from model \eqref{example 1 SDE}, define the random variables 
\begin{align}
\label{Y & Z sde}
    Y^{sde} &= \sum_{k=1}^n (y_{k-1}^{sde}-b)^2, \quad  Z^{sde} = \sum_{k=1}^n (y_k^{sde}-b) (y_{k-1}^{sde}-b).
\end{align}
Replacing the estimate $\hat{b}$ with the true $b$,  the estimator $\hat{\rho}^{sde}$ in \eqref{a_hat} is approximately 
\begin{align}\label{rho MLE approx}
    \hat{\rho}^{sde} \approx \frac{Z^{sde}}{Y^{sde}}.
\end{align}
The purpose of this approximation is to obtain tractable expressions so that moments of $\hat{\rho}^{sde}$ can be approximated, to evaluate bias as well as estimator variance. The estimate of the variance should be considered a lower bound, since assuming $b$ known leads to underestimating the variance, $V(\hat{\rho}^{sde})$. In Section \ref{Taylor expansion}, Taylor expansion approximations of mean and variance of a ratio of two random variables are given, and moments of $Z^{sde}$ and $Y^{sde}$ are calculated in Section \ref{sec:moments}. Substituting moments \eqref{E_Y sde}-\eqref{Cov_Y_Z sde} in eqs. \eqref{mean x1/x2}-\eqref{variance x1/x2} yield the following approximations of the expectation and variance of $\hat{\rho}^{sde}$ 
\begin{align}
    \label{Erho mle}
    &\begin{aligned}
    \frac{\mathbb{E}(\hat{\rho}^{sde})}{\rho} &\approx 
    1-\frac{2n(1-(\psi^2 -1)\rho^{2n-2})+\frac{1-\rho^{2n}}{1-\rho^2}2(\psi^2 -2)}{\left(n + \frac{1-\rho^{2n}}{1-\rho^2}(\psi^2 -1)\right)^2} ,    
    \end{aligned}\\
    \label{Vrho mle}
    &\begin{aligned}
    V(\hat{\rho}^{sde}) \approx & 
    \frac{1}{\left(n + \frac{1-\rho^{2n}}{1-\rho^2}(\psi^2 -1)\right)^2}   
    \left(
    (1-\rho(7+\rho+\rho^2)+4(\psi^2-1)\rho^{2n-1}(1+3\rho^2))\frac{n}{1+\rho} +     
    \right.\\
    &\left.
    ((\psi^2-1)(1-\rho(3+\rho+5\rho^2+8\rho^{2n})) -4\rho(2+\rho^2-\rho^{2n}))
    \frac{1-\rho^{2n}}{(1-\rho^2)(1+\rho)}
    \right)  
    \end{aligned}
\end{align}
where $\psi^2=2a(x_0 -b)^2/\sigma^2$. Since $(\mathbb{E}(\hat{\rho}^{sde})-\rho)$ and $V(\hat{\rho}^{sde})$ are of order $O(n^{-1})$, we have that $ \mathbb{E}(\hat{\rho}^{sde})/\rho \rightarrow 1 $ and $V(\hat{\rho}^{sde}) \rightarrow 0$ for $n\rightarrow \infty$, as also known from asymptotic theory since the MLEs are consistent. 


If we further have $\sigma^2/{2a} \ll (x_0 -b)^2$, i.e., $\psi^2 \gg 1$, it follows that 
$$ \frac{\mathbb{E}(\hat{\rho}^{sde})}{\rho} \approx 1 \quad \text{and} \quad V(\hat{\rho}^{sde}) \approx 0,  $$
implying that if the initial value is far from the mean compared to the size of the noise, the observed dynamics are approximately deterministic, and parameters are fully determined.

To derive approximate moments of $\hat{b}^{sde}$, we replace the estimate of $a$ with its true value in \eqref{b_hat}. Then (\cite{ditlevsen:2020}), 
\begin{align}
   \mathbb{E}(\hat{b}^{sde}) \approx b, \qquad
   V(\hat{b}^{sde}) \approx \frac{\sigma^2}{2a}\frac{1+\rho}{n(1-\rho)}.
\end{align}

If the data are generated from the ODE model \eqref{example 1 ODE}, we can still apply the SDE estimator. By substituting the data $\mathbf{y}^{ode}$ into equations \eqref{b_hat} - \eqref{a_hat}, we obtain estimators, which we will denote $\hat \rho^{ode\cdot sde}$ and $\hat b^{ode\cdot sde}$ to emphasize both the data-generating model and  the statistical model used for estimation. Note that when $x_0 \approx b$, the probability that the numerator of \eqref{a_hat} is negative can be large, which implies that $\hat{a}=-\frac{1}{\Delta}\log \hat{\rho}$ does not exist. 

Define the random variables $Y^{ode} = \sum_{k=1}^n (y^{ode}_k-b)^2$ and $Z^{ode} = \sum_{k=1}^n (y^{ode}_k-b) (y^{ode}_{k-1}-b)$ and substitute these into \eqref{mean x1/x2}-\eqref{variance x1/x2}. Using the moments of $Y^{ode}$ and $Z^{ode}$, calculated in Section \ref{sec:moments}, we approximate 
\begin{align}
    \label{E hat rho 0}
    &\begin{aligned}
        \frac{\mathbb{E}(\hat{\rho}^{ode\cdot sde})}{\rho}  = &  
        \frac{ \phi(n)}{\phi(n)+ n\sigma_0^2}  -  \frac{2 \sigma_0^2  \phi(n) \frac{2 -\rho^{2(n-1)} -\rho^{2n}}{1-\rho^{2n}}}{(\phi(n)+ n\sigma_0^2)^2}
         + \frac{\phi(n) (4\sigma_0^2 \phi(n) + 2n \sigma_0^4)}{(\phi(n)+ n\sigma_0^2)^3},
    \end{aligned}
    \\[1ex]
    \label{V hat rho}
    &\begin{aligned}
         V(\hat{\rho}^{ode\cdot sde}) = & \frac{ \sigma_0^2\phi(n)\frac{1+3\rho^2- 3\rho^{2n} -\rho^{2n+2}}{1-\rho^{2n}}+n\sigma_0^4}{(\phi(n)+ n\sigma_0^2)^2}
         -\frac{4\sigma_0^2 \phi(n)^2\frac{2\rho^2-\rho^{2n}-\rho^{2n+2}}{1-\rho^{2n}}}{(\phi(n)+ n\sigma_0^2)^3}
         + \frac{\rho^2\phi(n)^2 (4\sigma_0^2 \phi(n) + 2n \sigma_0^4)}{(\phi(n)+ n\sigma_0^2)^4},
    \end{aligned}
\end{align}
where $\phi(n) = (x_0-b)^2 \frac{1-\rho^{2n}}{1-\rho^2}$. When $n \rightarrow \infty$, both $\mathbb{E}(\hat{\rho}^{ode\cdot sde})/{\rho}$ and $V(\hat{\rho}^{ode\cdot sde})$ converge to 0, which means $\mathbb{E}(\hat{a}^{ode\cdot sde}) \rightarrow \infty$. This is because the observations in the ODE model are independent and thus, the autocorrelation is zero. The SDE estimator is thus correctly identifying the autocorrelation of the data, but is not able to identify the rate parameter. Therefore, for fixed $\Delta$ and large $n$, the bias in the parameter estimates will be large. However, perturbations introduce dependence in the data, which will help the SDE statistical model to identify also the rate parameter from data generated by an ODE model, as we will see below.

If $n$ is small, then $\rho^m \approx 1- m a\Delta$, $m \leq 2n$ and $\phi(n) \approx n (x_0 -b)^2 $. Thus, we have the following approximation
\begin{align*}
    &\begin{aligned}
        \frac{\mathbb{E}(\hat{\rho}^{ode\cdot sde})}{\rho}  \approx &  
        \frac{(x_0-b)^2}{(x_0-b)^2+ \sigma_0^2} \left( 1-  \frac{2\sigma_0^2}{n((x_0-b)^2+ \sigma_0^2)} + \frac{2\sigma_0^2(x_0-b)^2}{n((x_0-b)^2+ \sigma_0^2)^2}
         \right) + o(n^{-1}),
    \end{aligned}
    \\[1ex]
    &\begin{aligned}
     V(\hat{\rho}^{ode\cdot sde}) \approx & \frac{\sigma_0^2}{n((x_0-b)^2 + \sigma_0^2)} +
     \frac{(3-2a\Delta)\sigma_0^2(x_0-b)^2}{n((x_0-b)^2 + \sigma_0^2)^2} -
     \frac{4\sigma_0^2(x_0-b)^4}{n((x_0-b)^2 + \sigma_0^2)^3} - 
     \frac{(2-4a\Delta)\sigma_0^4(x_0-b)^2}{n((x_0-b)^2 + \sigma_0^2)^4}
    + o(n^{-1}).
    \end{aligned}
\end{align*}

If we furthermore have $\sigma_0^2 \ll (x_0 -b)^2$, then $\mathbb{E}(\hat{\rho}^{ode\cdot sde})/{\rho}$ and $V(\hat{\rho}^{ode\cdot sde})$ can be approximated by
$$\frac{\mathbb{E}(\hat{\rho}^{ode\cdot sde})}{\rho}  \approx 1 \quad \text{and} \quad V(\hat{\rho}^{ode\cdot sde}) \approx 0,$$
showing once more that if the initial value is far from the mean compared to the size of the noise, the observed dynamics are approximately deterministic, and parameters are fully determined.

If $n$ is large, then $\rho^{2(n-1)} \approx 0 $, $\rho^2 \approx 1-2a\Delta$ and $\phi(n) \approx (x_0 -b)^2/{2a\Delta} $. Then we have

$$\frac{\mathbb{E}(\hat{\rho}^{ode\cdot sde})}{\rho}  \approx  
        \frac{\frac{(x_0 -b)^2}{2a\Delta}}{\frac{(x_0 -b)^2}{2a\Delta}+n\sigma_0^2}  + o(n^{-1}) \quad \text{and} \quad V(\hat{\rho}^{ode\cdot sde}) \approx \frac{\sigma_0^2}{\frac{(x_0 -b)^2}{2a\Delta}+n\sigma_0^2} + o(n^{-1}) .$$

If furthermore $2a\Delta \sigma_0^2 \ll (x_0-b)^2$, then 
$$\frac{\mathbb{E}(\hat{\rho}^{ode\cdot sde})}{\rho}  \approx \frac{ \frac{(x_0 -b)^2}{2a\Delta}}{\frac{(x_0 -b)^2}{2a\Delta}+ n\sigma_0^2} \quad \text{and} \quad V(\hat{\rho}^{ode\cdot sde}) \approx 0.$$

As in \eqref{E hat rho 0}, the bias of $\hat{\rho}^{ode\cdot sde}$ increases for increasing sample size, due to the zero autocorrelation in the data. In Fig. \ref{fig: approximate bias MLE ODE} in the Appendix, we show how other factors, including $T$, $\Delta$, $(x_0-b)$ and $a$ influence $\mathbb{E}(\hat{\rho}^{ode\cdot sde})/\rho$. Smaller $a$ and larger $x_0-b$ yield $\mathbb{E}(\hat{\rho}^{ode\cdot sde})/{\rho}$ closer to 1, while for fixed $a$ and $x_0-b$, shorter $T$ and larger $\Delta$ lead to better estimates.

We approximate the moments of $\hat{b}^{ode\cdot sde}$ with $a$ at its true value by
\begin{align}
    & \mathbb{E} (\hat{b}^{ode\cdot sde}) \approx b, \quad \mbox{V} (\hat{b}^{ode\cdot sde}) \approx \left ( \frac{n-1}{n^2}+\frac{1+\rho^2}{n^2(1-\rho)^2} \right ) \sigma_0^2.
\end{align}
Larger sample size $n$ yields smaller variance of $\hat{b}^{ode\cdot sde}$, as expected.

We now consider different types of perturbation to the dynamics. The perturbations only happen in the data-generating model and are assumed unobserved and unknown, such that the previous estimators, assuming no perturbations, are used for parameter estimation. We then analytically calculate the approximate bias induced by the perturbations. 

\paragraph{Instantaneous perturbation}\label{Instantaneous perturbation.}

Assume that at time $t_p$, where $t_0 < t_p < t_n$, an instantaneous shift  of size $h$ of the state occurs, so that the value of $x_{t_p}$ jumps to $x_{t_p}+h$. Thus, the true model takes the form
\begin{align}
\label{model instantaneous}
\begin{array}{ll}
dx_t = -a (x_t - b)dt + \sigma dW_t , &  t \neq t_p,\\
\hspace{2mm} x_t = x_{t-} +h, &  t = t_p,
\end{array}
\end{align}
where $\sigma = 0$ in the ODE model. In Figure \ref{fig: example 1 trajectory comparison}, an example trajectory with $h = 0.5$, $t_p = 50$ and $T=100$ is shown.

With $\mathbf{y}^{ode}$ generated from \eqref{model instantaneous}, we assume the ODE model \eqref{example 1 ODE} and denote the estimators $\hat{\rho}^{ode}_1$ and $\hat{b}^{ode}_1$, or we assume the SDE model \eqref{example 1 SDE} and denote the estimators $\hat{\rho}^{ode\cdot sde}_1$ and $\hat{b}^{ode\cdot sde}_1$, where the subscript $1$ indicates that the estimator is fitted to perturbed data. Likewise, with data $\mathbf{y}^{sde}$ generated from \eqref{model instantaneous}, we denote estimators $\hat{\rho}^{sde}_1$ and $\hat{b}_1^{sde}$, assuming the SDE model \eqref{example 1 SDE}, or $\hat{\rho}^{sde\cdot ode}_1$ and $\hat{b}_1^{sde\cdot ode}$, assuming the ODE model \eqref{example 1 ODE}.  

Define $Y^{ode}_1$ and $Z^{ode}_1$ as above, but for the perturbed data. Using their moments calculated in Section \ref{Calculation ode instantaneous}, we approximate the expectation of $\hat{\rho}^{ode\cdot sde}_1$ up to second order in $h$,
\begin{align}
\label{E rho_1 ode approx}
\mathbb{E}(\hat{\rho}^{ode\cdot sde}_1) &\approx \mathbb{E}(\hat{\rho}^{ode\cdot sde}) +  A_1(\rho, n,k, \sigma_0^2)h + A_2(\rho, n,k, \sigma_0^2) h^2 + o(h^2),
\end{align}
where $\mathbb{E}(\hat{\rho}^{ode\cdot sde})$ is given in \eqref{E hat rho 0}, $A_1(\rho, n, k, \sigma_0^2)$ and $A_2(\rho, n, k, \sigma_0^2)$ are given in Section \ref{Calculation ode instantaneous}, and $k$ is the smallest integer such that $t_p \leq t_k$. Not only does the size of the perturbation $h$ affect the bias, but also when the perturbation occurs in the observation interval, indicated by the observation index $k$. Both $A_1(\rho, n, k, \sigma_0^2)$ and $A_2(\rho, n, k, \sigma_0^2)$ are of order $n^{-1}$, so the impact of the perturbation decreases with increasing $n$. However, since $\mathbb{E}(\hat{\rho}^{ode\cdot sde})$ converges to 0, also $\mathbb{E}(\hat{\rho}^{ode\cdot sde}_1) \rightarrow 0$ as $n \rightarrow \infty$.

We then turn to the situation where $n$ is finite. If the perturbation occurs late, i.e.,  $t_p \approx T$ and $k = n$, the impact of the last observation on the expectation of $\hat{\rho}^{ode\cdot sde}_1$ is approximately 
\begin{align*}
    \mathbb{E}(\hat{\rho}^{ode\cdot sde}_1) = \mathbb{E}(\hat{\rho}^{ode\cdot sde}) + \frac{x_0-b}{\mathbb{E}(Y_0)}
    \rho^{k-1}
    \left(
    1-\frac{2\sigma_0^2}{\mathbb{E}(Y_0)}+\frac{V(Y_0)}{(\mathbb{E}(Y_0))^2}
    \right)h.
\end{align*}
We furthermore have
\begin{align}
    \mathbb{E}(\hat{b}^{ode\cdot sde}_1) \approx b+\frac{e^{-a(t_k-t_p)}}{n(1-\rho)}h.
\end{align}
For fixed $t_p$, the bias of $\hat{b}^{ode\cdot sde}_1$ is still of order $n^{-1}$. Thus, $\mathbb{E}(\hat{b}^{ode}_1) -b \approx 0$ as $n \rightarrow \infty$.

We approximate $\hat{\rho}^{sde}_1$ and $\hat{b}^{sde}_1$ in the same way. Define $Y^{sde}_1 $ and $Z^{sde}_1 $ as above but for perturbed data generated from the SDE model \eqref{model instantaneous}, moments can be found in Section \ref{sec:moments}. The expressions for $\mathbb{E}(\hat{\rho}^{sde}_1)$ are complex and therefore not given. However, from the moments of $Y^{sde}_1$ and $Z^{sde}_1$ in eqs.\eqref{EY sde 1}-\eqref{CovYZ sde 1} it follows that the coefficients of $h$ and $h^2$ in the expression for $\mathbb{E}(\hat{\rho}^{sde}_1)$ still are of order $n^{-1}$, so that $\mathbb{E}(\hat{\rho}^{sde}_1) \rightarrow \mathbb{E}(\hat{\rho}^{sde})$ as $n \rightarrow \infty$.

If $t_p \approx T$ and $k=n$, then the impact of the last observation is 
\begin{align*}
    \mathbb{E}(\hat{\rho}^{sde}_1) \approx \mathbb{E}(\hat{\rho}^{sde}) + \frac{x_0-b}{\mathbb{E}(Y)}
    \rho^{k-1}
    \left(
    1-\frac{\sigma^2}{a\mathbb{E}(Y)}
    \left(
    k-\frac{1-\rho^{2k}}{1-\rho^2}
    \right)
    +\frac{V(Y)}{(\mathbb{E}(Y))^2}
    \right)h.
\end{align*}

We easily obtain that $\mathbb{E}(\hat{b}^{ode\cdot sde}_1)=\mathbb{E}(\hat{b}^{sde}_1)$, however, this is assuming $a$ known. 
The estimators $\hat{b}^{ode\cdot sde}_1$ and $\hat{b}^{sde}_1$ are related to $\hat{\rho}^{ode\cdot sde}_1$ and $\hat{\rho}^{sde}_1$. Since $\hat{\rho}^{ode\cdot sde}$ has a non-negligible bias in most conditions (see Section \ref{Estimators Example 1}), it is reasonable to expect $\hat{\rho}^{ode\cdot sde}_1$ to be biased as well.

\paragraph{Randomly varying long-term mean.}
\label{The oscillation of the long-term mean}
Perturbations can also be caused by time-varying parameters. In this section, we consider the case where the long-term mean is not a fixed value but disturbed by white noise. 

For the models \eqref{example 1 ODE} and \eqref{example 1 SDE}, we still assume $a$ constant, but let the long-term mean $b_t$ follow a normal distribution, $b_t \sim \mathcal{N}(b,\sigma_b^2)$, and $b_t$ is independent of $b_s$ for all $s \neq t$. 
An example trajectory with $b_t \sim \mathcal{N}(0,1)$ is shown in Figure \ref{fig: example 1 trajectory comparison}. 
The conditional expectation of $x_{k+1}$ given $x_k$ are the same for both the ODE and SDE models,
$$E(x_{k+1}|x_k) = \rho x_k + b(1-\rho),$$
which is the same as in the original models \eqref{example 1 ODE} and \eqref{example 1 SDE} (for the original ODE model, it is not an expectation but the solution). However, the randomness of $b_t$ increases the conditional variance of $x_{k+1}|x_k$ by $(1-\rho)^2\sigma_b^2$, 
\begin{align}
    \label{con variance ode}
    &\mbox{V}(x_{k+1}^{ode}|x_k^{ode}) = \sigma_b^2(1-\rho)^2,\\
    \label{con variance sde}
    &\mbox{V}(x_{k+1}^{sde}|x_k^{sde}) =\left ( \frac{\sigma^2}{2a} + \sigma_b^2\right ) (1-\rho)^2.
\end{align} 
Thus, we expect larger variances of parameter estimates. 

The rate $a$ affects the variances through $\rho$. From \eqref{con variance ode} it follows that the variance increases with $a$ in the ODE model, while for the SDE model, the change is not monotone. However, we obtain the following condition: As long as $a$ satisfies
$$\frac{1-(1+2a\Delta)\rho^2}{2a\Delta\rho(1-\rho)} > \frac{\sigma_b^2}{\sigma^2/2a},$$
then larger $a$ leads to smaller variance; otherwise it leads to larger variance. 

\paragraph{Error term in differential equations.}
\label{Error term in differential equations}
The drift function $\eqref{drift example 1}$ might also be misspecified. In this section, we consider a small error term $g(x) = O(x-b)^2$ such that the drift function changes from $-a(x_{t}-b)$ to $-a(x_{t}-b) + g(x_{t})$. Let $g(x)$ be of the form $g(x) = \gamma_t (x-b)^2$ where $\gamma_t \sim \mathcal{N}(0,\sigma_{\gamma}^2)$ and $ \gamma_t$ is independent of $\gamma_s$ for all $s \neq t$. Thus, the model becomes
\begin{align}
    \label{error term}
    &dx_{t} = -a(x_{t}-b)dt +  \gamma_t (x-b)^2 dt + \sigma dW_t,
\end{align}
where, as usual, $\sigma = 0$ for the ODE model. An example trajectory with $\gamma_t \sim \mathcal{N}(0,0.04)$ is shown in Fig. \ref{fig: example 1 trajectory comparison}. For small $\Delta$ we have $E(x_{t+\Delta} -x_{t} | x_t) \approx  \left(-a(x_t -b) + \gamma (x_t -b)^2 \right) \Delta $ for both the ODE and the SDE model. 

In Section \ref{sim: Error term in differential equations}, we will study how this type of error term affects the parameter estimation in a simulation study.

\subsection{Example 2: Perturbations and simplifications in  epidemic models}\label{perturbaion Example 2}

\paragraph{Occurrence of unexpected events.}
Unexpected events are common in the spread of epidemics. For example, a large gathering of people or the introduction of a powerful medicine may result in an instantaneous change in the epidemiological status of the population without changing the mechanism of transmission of the disease. We assume that the spread of the epidemic still follows the SIR model \eqref{model SIR}. However, at time $t_p$, an instantaneous change occurs that results in a change of a certain level in the population sizes of S and/or I. Since both $s_t$ and $i_t$ are assumed continuous in $t \in (t_0,t_p)$, the perturbation is represented as
\begin{align}
\label{SIR perturbation}
    \begin{pmatrix}
        s_{t_p}\\
        i_{t_p}
    \end{pmatrix} = \lim\limits_{t \uparrow t_p} 
    \begin{pmatrix}
        s_{t}\\
        i_{t}
    \end{pmatrix}
    + \begin{pmatrix}
        h_1\\
        h_2
    \end{pmatrix}.
\end{align}
Constants $h_1$ and $h_2$ should be such that $s_{t_p}$ and $i_{t_p}$ remain within their domains, i.e., $s_{t_p}+i_{t_p}\leq 1$. An example of a perturbation is if a large part of the population is vaccinated at a certain point in time, then people from compartment $S$ will be moved to compartment $R$, which can be modeled by setting $h_1 < 0, h_2=0$. If there is a large outbreak of mass infection due to a single large gathering (superspread event), then people from compartment $S$ will be moved to compartment $I$, which can be modeled by setting $-h_1 = h_2>0$. The perturbation is assumed not to affect the contact and removal rates. 

We show an example of the effect of the perturbation in the third column of Fig. \ref{fig: SIR model}, which can be compared with the case without perturbation in the first column. We set $t_p = 21$ and $(h_1,h_2)^\top = (-0.02,0.02)^\top$, which models a superspread event, where $2\%$ of the total population in the susceptible population becomes infectious at time $t_p$. We set $T=40$, $\Delta = 0.5$, $\alpha = 0.5$, $ \beta = 0.3$, $ \sigma_1 = 5\times10^{-3}$, $ \sigma_2 = 1\times10^{-3}$, $\gamma_1 = \sigma_1\sqrt{T}$, $\gamma_2 = \sigma_2/\sqrt{2(\beta - \alpha s^*)}$ and initial values $s_0 = 0.99, i_0 = 0.001$.

Although the parameter values remain the same, the instantaneous shift in the state variables affects the subsequent dynamics, as well as the estimated parameters. We will explore the effect of perturbations on the parameter estimates of the ODE and SDE models.

\paragraph{SEIR model.}\label{sec:SEIR model}
The SIR model can be simplified (e.g., SI and SIS) or extended (e.g., SEIR and MSIR). The SEIR model \cite{aron:1984} is a popular modification of the SIR model, which adds an intermediate exposure stage when a susceptible individual becomes infected but is not yet infectious. The SDE version of the SEIR model is 
\begin{eqnarray}
\label{SEIR model}
    ds_t &=& -\alpha s_ti_t dt + \sigma_1 dW_t^1, \nonumber  \\
    de_t &=& (\alpha s_ti_t - \lambda e_t)dt + \sigma_2 dW_t^2,\\
    di_t &=& (\lambda e_t - \beta i_t)dt + \sigma_3 dW_t^3,\nonumber 
\end{eqnarray}
where $\lambda >0$ is the transition rate from the exposed state to the infectious state, and $(W_t^1, W_t^2, W_t^3)$ are independent standard Wiener processes. The ODE model is obtained by setting $\sigma_1 = \sigma_2 = \sigma_3 = 0$. In the ODE model, the observations have independent measurement errors as in \eqref{SIR deterministic error}, with variances $(\gamma_1^2, \gamma_2^2, \gamma_3^2)$. In Figure \ref{fig: SIR model}, we show ODE and SDE example trajectories of the SEIR model with $T = 40$ and $\Delta =0.5$, $\alpha = 0.5$, $\lambda =1$, $ \beta = 0.3$, $ \sigma_1 = 5\times10^{-3}$, $ \sigma_2 =\sigma_3 = 1\times10^{-3}$, $ \gamma_1 = \sigma_1\sqrt{T}$, $ \gamma_2 = \sigma_2/\sqrt{2\lambda}$, and $\gamma_3 =\sigma_3/\sqrt{2\beta}$ and initial values $s_0 = 0.99, e_0 = i_0 = 0.001$.

By including the extra exposure compartment, the SEIR model reflects better the transmission patterns of some infectious diseases, such as COVID-19, than the SIR model. However, usually it is not possible to distinguish between individuals in the susceptible and the exposed compartments from data, and statistical inference therefore simplifies when assuming a SIR model, where all compartments are observed. Here, we investigate how model misspecification affects parameter estimation, by assuming the simpler SIR model when data are generated from the SEIR model. 

We fit the SIR model using observations from the SEIR model $\{(s_k, p_k), k=1,\ldots, n\}$ to estimate the contact rate $\alpha$ and removal rate $\beta$, where we set $p_k = e_k + i_k$.

The SEIR model \eqref{SEIR model} can be rewritten with state variables $(s_t, p_t)$, which in the ODE version is 
\begin{align}
\label{approximate SEIR}
\begin{aligned}
    ds_t &= -\alpha s_t p_t dt + \alpha s_t e_t dt,\\
    dp_t &= (\alpha s_t p_t - \beta p_t)dt - (\alpha s_t e_t - \beta e_t) dt.
\end{aligned}
\end{align}
The equations for $(s_t, p_t)$ correspond to the SIR model, except for the additive terms $\alpha s_t e_t$ and $\alpha s_t e_t - \lambda e_t$, which contain $e_t$. Thus, the difference between the SIR and SEIR models will be small when $e_t$ is small.

We have no analytical solution of the SEIR model, but we can represent $e_t$ and $i_t$ in terms of $s_t$ (see \cite{Heng:2020}) 
\begin{align}
\begin{aligned}
    e_t &= 1- i_0 -\frac{\lambda}{\lambda+\beta}s_0 - \frac{\beta}{\lambda+\beta}\left(s_t -  \frac{\beta}{\alpha}\ln\frac{s_t}{s_0}\right), \\
    i_t &= i_0 -\frac{\lambda}{\lambda+\beta}(s_t-s_0)+\frac{\beta}{\alpha} \frac{\lambda}{\lambda+\beta} \ln\frac{s_t}{s_0}.
\end{aligned}
\end{align}
This yields $p_t = 1-s_t +\frac{\beta}{\alpha}  \ln\frac{s_t}{s_0}$. If $i_0 \approx \frac{\lambda}{\lambda + \beta} (1-s_0)$, then $\frac{i_t}{p_t} \approx \frac{\lambda}{\lambda + \beta}$, and the approximated model \eqref{approximate SEIR} becomes
\begin{align}
\begin{aligned}
    ds_t &= - \alpha's_t p_t dt,\\
    dp_t &= (\alpha' s_t p_t - \beta' p_t)dt,
\end{aligned}
\end{align}
with $\alpha' = \frac{\lambda\alpha}{\lambda + \beta}$ and $\beta' = \frac{\lambda\beta}{\lambda + \beta}$.

For $\lambda \gg \alpha, \beta$, the incubation period in compartment E is much shorter compared to the time spent in compartment I, leading to $\lambda e_t \rightarrow \alpha (1-i_t-r_t)$, and the dynamics are close to a SIR model (see \cite{Kozyreff:2022}). 

\section{Simulations}\label{Simulations}
We now perform a simulation study of the models in Section \ref{Example 1} and \ref{Example 2}, estimating parameters by the methods in Section \ref{Estimators}. To minimize the sum of squared errors \eqref{lse formula}, we use \texttt{optim} in \texttt{R} \cite{R} with the  Brodyden-Fletcher-Goldfarb-Shanno (BFGS) method \cite{Nocedal:2006}. 

We fit the ODE and the SDE models to data generated both from ODE and SDE models. We denote by ODE-ODE and SDE-SDE the estimates from correctly specified models, and ODE-SDE and SDE-ODE the estimates from misspecified models, with notation (data-generating model)-(model used for estimation). Thus, for each scenario, we have four sets of parameter estimates, of which two sets are from misspecified models. Furthermore, we conduct simulation studies of the perturbation scenarios discussed in Section \ref{Model misspecification}. For the perturbations, we use the same set of random numbers to generate trajectories as used for simulating the original model, so the differences are due only to the perturbations, not to random variation. 

\subsection{Example 1: Estimation in linear model with perturbations}
Here we investigate the linear model \eqref{example 1 ODE} and \eqref{example 1 SDE} using estimators \eqref{lse formula} (assuming an ODE model) or \eqref{b_hat} - \eqref{a_hat} (assuming an SDE model). 

\paragraph{Without perturbation.}
\label{sim: basic models}
We consider three observation schemes: (1) A long observation interval, $T=100$, and a large time step, $\Delta = 2$, so that $n=50$; (2) A short observation interval, $T=50$, and a small time step, $\Delta = 1$, so again $n=50$; (3) A short observation interval, $T=50$, and a large time step, $\Delta = 2$, so that $n=25$. The rate parameter $a$ is set to $0.05$ or $0.1$. The remaining parameters are $b=0$, $\sigma/\sqrt{2a} = \sigma_0 = 0.05$ and the initial value is set to $x_0 = 5$.

For each scenario, 10,000 trajectories of models \eqref{example 1 ODE} and \eqref{example 1 SDE} were simulated. Four sets of estimates were computed from each of the data sets. In Appendix, the density plots of the 10,000 estimates Fig. \ref{fig: basic model simulation 1} and Fig. \ref{fig: basic model simulation 2} provide $\hat{\rho}/\rho$ and $\hat{b}$ for $a = 0.05$ and $a = 0.1$, respectively.

Fitting the correct model (\textsc{ODE-ODE} and \textsc{SDE-SDE}) perform well because of the simple linear form of the model and no misspecification of the noise. The misspecified SDE-ODE also leads to good estimates due to the model being linear. However, for the ODE-SDE case, $\rho$ is underestimated and $\hat{\rho}/{\rho}$ is closer to one for smaller $T$ and larger $\Delta$, which is in agreement with the theoretical results in Section \ref{Estimators Example 1 no perturbation}. 
A larger $T$ improves the estimate of $b$, as does a larger $a$, also in agreement with the theoretical results.

\paragraph{Instantaneous perturbation.}
\label{sim: Instantaneous perturbation}
\begin{figure}[ht]
    \centering
    \includegraphics[width = 0.8\textwidth]{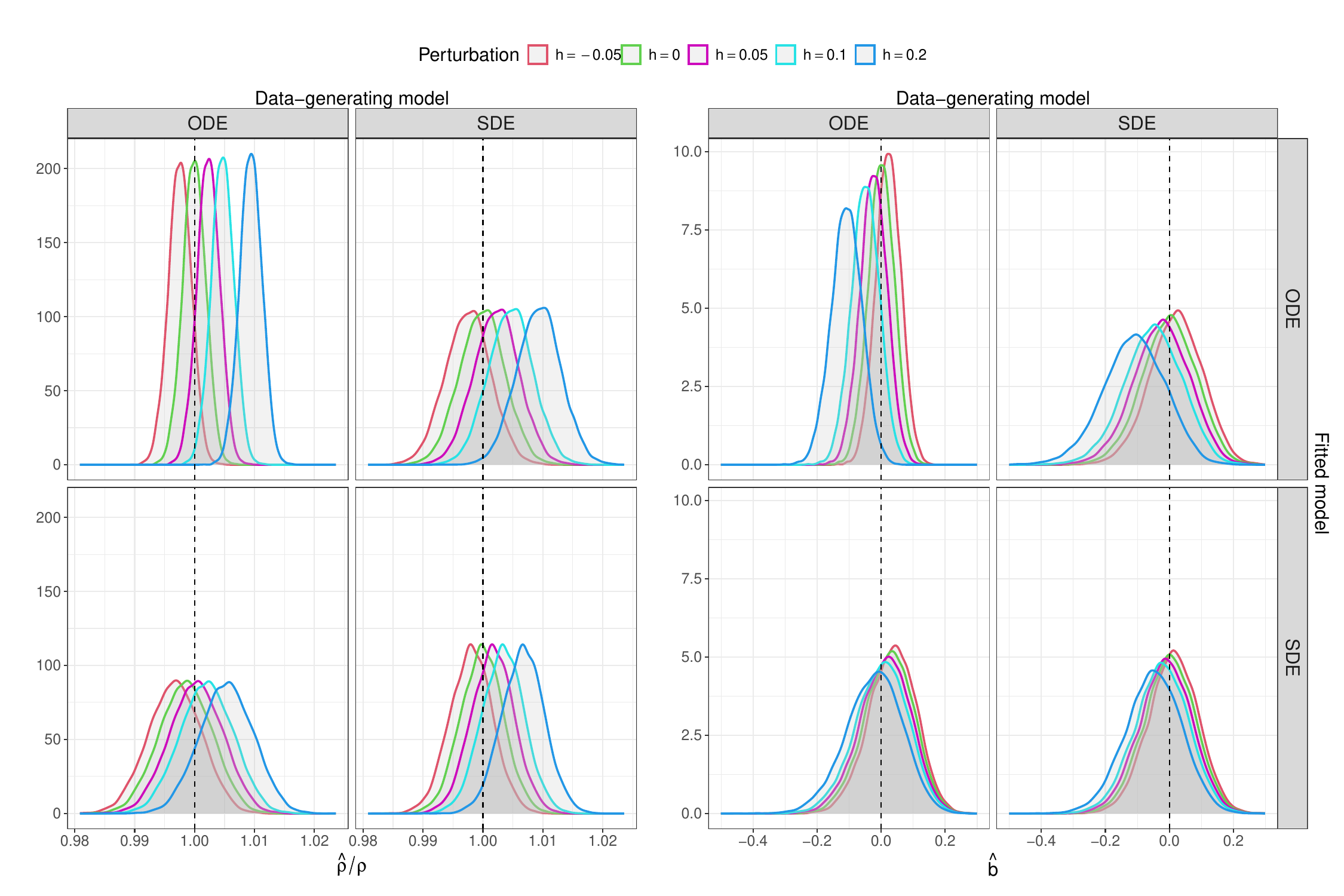} 
    \caption{{\bf Linear model with instantaneous perturbation.} Density plots of parameter estimates based on 10,000 simulated data sets of models \eqref{example 1 ODE} and \eqref{example 1 SDE} with instantaneous perturbation \eqref{model instantaneous} with $T=50$ and different values of $h$. Using data from \texttt{Data-generating model} (columns), we fit it to the \texttt{Fitted model} (rows). The black dashed lines indicate the true values. Parameters are $a= 0.05, b=0$, $\sigma_0 = 0.05$ and $\sigma = \sigma_0 \sqrt{2a}$. }
    \label{fig: instantaneous perturbation sim1}
\end{figure}

We simulated trajectories of the ODE and SDE models with instantaneous perturbation, see Section \ref{Estimators Example 1}, with $\Delta =2$ and either $T= 50$ or $100$. We set the perturbation to happen at $t_p = 10$ and explore different perturbation sizes $h \in \{-0.2, -0.1, 0, 0.1, 0.2\}$. Parameters are $a = 0.05$, $b=0$, $\sigma_0 = 0.05$ and $\sigma^2 = 2a \sigma_0^2$. For each scenario, we simulated 10,000 data sets, and estimates were computed using \eqref{lse formula} and \eqref{b_hat}-\eqref{a_hat}, respectively. Density plots of the estimates $\hat{\rho}/{\rho}$ and $\hat{b}$ are shown in Fig. \ref{fig: instantaneous perturbation sim1} for $T=50$ and in Fig. \ref{fig: instantaneous perturbation sim2} in Appendix for $T=100$. We clearly see that using an SDE model for estimation removes or attenuates the bias induced by the perturbation, whatever the data-generating model. We conclude that even for such a small perturbation, the SDE model helps regularizing the estimation problem, whereas the ODE model cannot adapt. For larger $T =100$, the instantaneous perturbation has less effect, as expected. However, if events of instantaneous perturbations occur at a certain frequency, longer observation intervals would likely not remove the bias.

\paragraph{Randomly varying long-term mean.}
\label{sim： The oscillation of the long-term mean }

We simulated trajectories of the ODE and SDE models with randomly varying long-term mean, see Section \ref{The oscillation of the long-term mean}, using $T = 100$ and $\Delta =2$. At each time point $t$, we draw a new $b_t$ from a normal distribution $\mathcal{N}(b, \sigma_b^2)$ with $\sigma_b \in \{0.5, 1, 2\}$. We set either $a=0.05$ or $0.1$. Other parameters are $b=0$, $\sigma_0 = 0.05$ and $\sigma^2 = 2a \sigma_0^2$.

\begin{table}[ht]
\centering
\begin{tabular}{ccc cc cc}
\toprule
\multirow{2}{*}{$\sigma_b$} 
& \multirow{2}{*}{\makecell{Data-generating\\model}} 
& \multirow{2}{*}{\makecell{Fitted\\model}} 
& \multicolumn{2}{c}{Variance of $\hat{\rho} / \hat{\rho}_0$} 
& \multicolumn{2}{c}{Variance of $\hat{b}- \hat{b}_0$} \\
\cmidrule(lr){4-5} \cmidrule(lr){6-7}
& & & $a=0.05$ & $a=0.1$ & $a=0.05$ & $a=0.1$ \\
\midrule
\multirow{4}{*}{0.5} 
  & \multirow{2}{*}{ODE} & ODE & 1.97e-7 & 1.26e-6 & 4.54e-5 & 3.33e-5 \\
  &                      & SDE & 1.71e-7 & 1.04e-6 & 4.05e-5 & 3.12e-5 \\
  \cline{2-7}
  & \multirow{2}{*}{SDE} & ODE & 1.95e-7 & 1.26e-6 & 4.53e-5 & 3.33e-5 \\
  &                      & SDE & 1.67e-7 & 1.01e-6 & 4.08e-5 & 3.13e-5 \\
\midrule
\multirow{4}{*}{1} 
  & \multirow{2}{*}{ODE} & ODE & 7.78e-7 & 5.04e-6 & 1.81e-4 & 1.33e-4 \\
  &                      & SDE & 6.83e-7 & 4.14e-6 & 1.62e-4 & 1.25e-4 \\
    \cline{2-7}
  & \multirow{2}{*}{SDE} & ODE & 7.78e-7 & 5.06e-6 & 1.81e-4 & 1.33e-4 \\
  &                      & SDE & 6.67e-7 & 4.06e-6 & 1.63e-4 & 1.25e-4 \\
\midrule
\multirow{4}{*}{2} 
  & \multirow{2}{*}{ODE} & ODE & 3.11e-6 & 2.02e-5 & 7.26e-4 & 5.33e-4 \\
  &                      & SDE & 2.73e-6 & 1.66e-5 & 6.47e-4 & 4.99e-4 \\
    \cline{2-7}
  & \multirow{2}{*}{SDE} & ODE & 3.11e-6 & 2.02e-5 & 7.25e-4 & 5.33e-4 \\
  &                      & SDE & 2.67e-6 & 1.62e-5 & 6.53e-4 & 5.01e-4 \\
\bottomrule
\end{tabular}
\caption{{\bf Linear model with randomly varying long-term mean.} Variances of the parameter estimates $\hat{\rho} / \hat{\rho}_0$ and $\hat{b} - \hat{b}_0$ for $\sigma_b \in \{0.5, 1, 2\}$ and $a \in \{0.05, 0.1\}$ based on 10,000 simulated data sets of models \eqref{example 1 ODE} and \eqref{example 1 SDE} with randomly varying long-term mean \eqref{The oscillation of the long-term mean} with $T=100$, $\Delta = 2$. \label{tab:variance-comparison}}
\end{table}

For each scenario, we simulated 10,000 data sets. To visualize the effect of a random $b_t$, we compare estimates $(\hat \rho, \hat b)$ with those obtained from the data set without perturbation $(\hat\rho_0, \hat b_0)$, for the same set of random numbers used in the simulation of each data set. Density plots of the ratio $\hat{\rho} / \hat{\rho}_0$ and the difference $\hat{b} - \hat{b}_0$ are shown in Figure \ref{fig: oscillation simulation 1} for $a = 0.05$ and in Figure \ref{fig: oscillation simulation 2} in the Appendix for $a = 0.1$. 

Since the perturbation is symmetric around the non-perturbed $b$, the perturbation does not induce any bias, but increases the variance of the estimators of both $\rho$ and $b$. As expected, the variance increases with increasing $\sigma_b$. Interestingly, also a larger $a$ increases the variance induced by the perturbation in $b$. The estimator variance induced by the perturbation is always smaller when using the SDE model \eqref{example 1 SDE} for model fitting, regardless of the data-generating model, see Table \ref{tab:variance-comparison}. The reduction in variance is up to 20\%, once again illustrating the advantage of assuming an SDE model over an ODE model.

\paragraph{Error term in differential equations.}
\label{sim: Error term in differential equations}
We simulated 10,000 trajectories of the ODE and SDE models with a squared error term in the differential equations, see Section \ref{Error term in differential equations}, using $T = 100$ and $\Delta =2$. We drew the coefficient of the error term $\gamma_t$ from a normal distribution $\mathcal{N}(0,\sigma_{\gamma}^2)$ independently and used the same $\gamma_t$ in the ODE and SDE models at each time step $t$. We used different values of the variance of $\gamma_t$,  $\sigma_{\gamma} \in \{0.005, \, 0.01, \, 0.015\}$, of $a=0.01$ or $a = 0.02$ and different initial conditions $x_0= 5$ or $x_0 = 10$. Other parameters are $b=0$, $\sigma_0 = 0.05$ and $\sigma^2 = 2a \sigma_0^2$. 

\begin{figure}[ht]
    \centering
    \includegraphics[width = 0.8\textwidth]{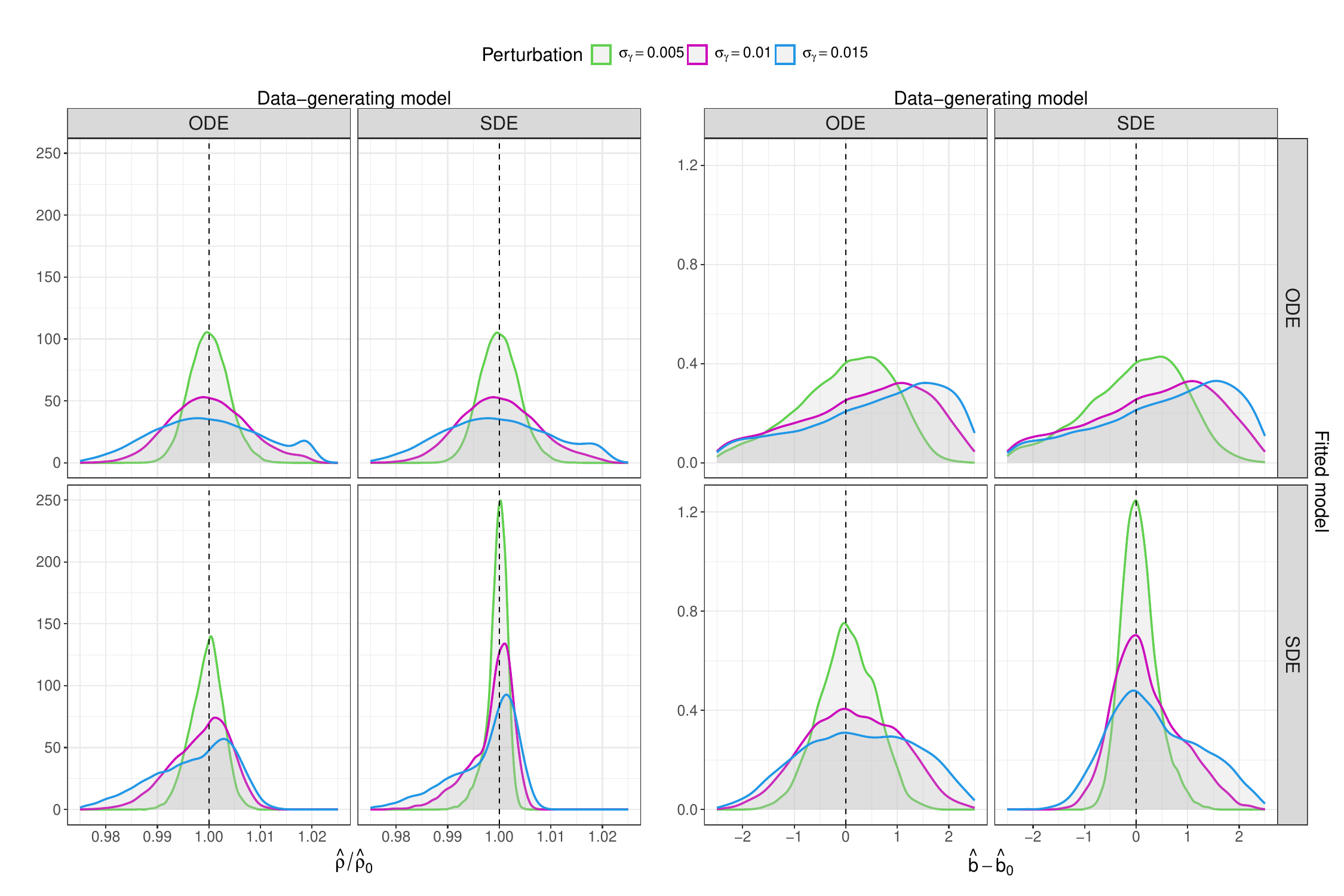} 
    \caption{{\bf Linear model with quadratic misspecification term.} Density plots of parameter estimates from 10,000 simulated data sets. Models with squared error term $\gamma_t (x_t-b)^2$ were simulated from models \eqref{example 1 ODE} (ODE) or \eqref{example 1 SDE} (SDE). Data from the \texttt{Data-generating model} (columns), were fitted to the \texttt{Fitted model} (rows). The black dashed lines indicate the true values. Parameters are $a = 0.01, x_0 = 10$, $\sigma_0 = 0.05$ and $\sigma = \sigma_0 \sqrt{2a}$.}
    \label{fig: Error term in differential equations sim1}
\end{figure}

To visualise the effect of a quadratic error term in the drift, we compare estimates $(\hat \rho, \hat b)$, to those obtained from the data sets with no perturbation, $(\rho_0, b_0)$, for the same set of random numbers used in the simulation of each data set. We again compute the ratio $\hat{\rho} / \hat{\rho}_0$ and the difference $\hat{b} - \hat{b}_0$. The density plots of $\hat{\rho} / \hat{\rho}_0$ and $\hat{b} - \hat{b}_0$ for different $\sigma_{\gamma}$'s can be found in Figure \ref{fig: Error term in differential equations sim1} $(a = 0.01, x_0 = 10)$, \ref{fig: Error term in differential equations sim2} $(a = 0.01, x_0 = 5)$ and \ref{fig: Error term in differential equations sim3} $(a = 0.02, x_0 = 10)$.

Not surprisingly, a larger variance of $\gamma_t$ has a larger impact on the parameter estimates. For smaller $a$ or larger initial value $x_0$, the perturbation term has a greater impact on the parameter estimates, agreeing with the results in Section \ref{Error term in differential equations}. Regardless of whether the data are generated from an ODE or an SDE model, when $a$ is small, assuming an SDE model for parameter estimation greatly reduces the negative effect of the perturbation term. For larger $a$, the advantage of the SDE model is less clear. 

\subsection{Example 2: Estimation in perturbed and simplified epidemic models}\label{SIR model sim}
Here, we investigate the epidemic models of Sections \ref{Example 2} and \ref{perturbaion Example 2}. The models are computationally heavy, so only 1,000 data sets were simulated for each scenario. Throughout we use $\Delta=0.5$ and $T=40$. 

\paragraph{SIR model.}\label{sim SIR model full dataset}

\begin{figure}[ht]
    \centering
    \includegraphics[width = 0.8\textwidth]{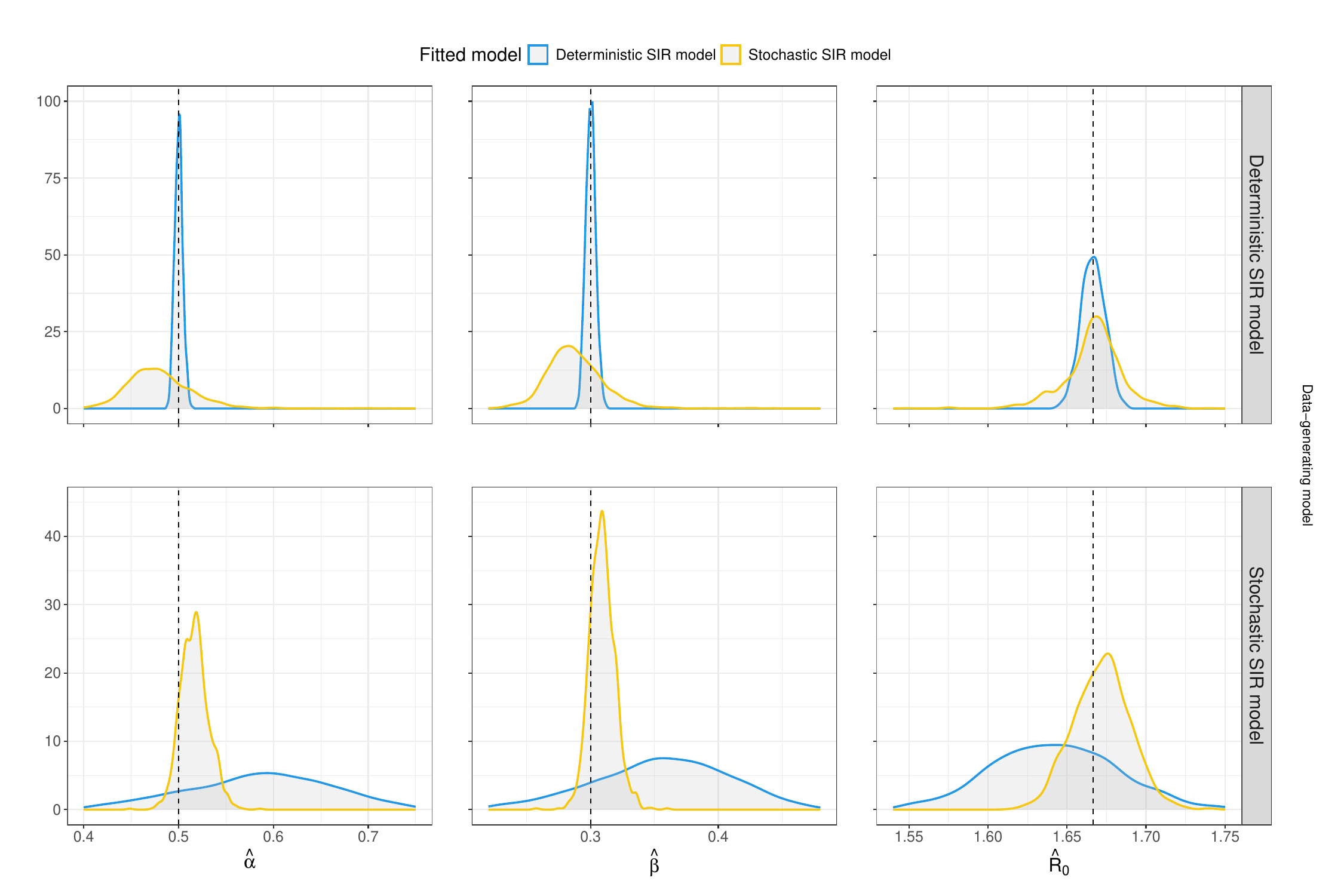} 
    \caption{{\bf SIR model without perturbation}. Density plots of 1,000 estimates of the contact rate $\alpha$ and the removal rate $\beta$ in the deterministic and stochastic SIR models \eqref{model SIR}. Data were simulated using $\Delta = 0.5$, $T = 40$, $\alpha = 0.5$, $ \beta = 0.3$, $ \sigma_1 = 3\times10^{-3}$, $ \sigma_2 = 1\times10^{-3}$, $\gamma_1 = \sigma_1\sqrt{T}$, $ \gamma_2 = \sigma_2/\sqrt{2(\beta - \alpha s^*)}$, and initial values $s_0 = 0.99, i_0 = 0.001$. Dashed lines are at the true values used in the simulation.
    }
    \label{fig: SIR original data set}
\end{figure}
\begin{table}[ht]
\centering
    \begin{tabular}{c c c c c} 
    \hline
    & \multicolumn{4}{c}{Data-generating model}\\
       & \multicolumn{2}{c}{Deterministic SIR }  & \multicolumn{2}{c}{Stochastic SIR}  \\
     Fitted model & $\hat{\alpha}$ & $\hat{\beta} $ & $\hat{\alpha}$ & $\hat{\beta} $ \\
     \hline\hline
    \multirow{2}{9em}{Deterministic SIR}  & 0.500 &  0.300 & 0.585 & 0.360\\
      &(1.78$\times10^{-5}$) & (1.46$\times10^{-5}$)  & (6.19$\times10^{-3}$)& 2.80$\times10^{-3}$)\\
      \hline
     \multirow{2}{9em}{Stochastic SIR}  & 0.480 & 0.287 & 0.510 & 0.309\\
     &(1.33$\times10^{-4}$) &(5.38$\times10^{-4}$) & (2.18$\times10^{-4}$ )& (9.71$\times10^{-4}$)\\
     \hline
    \end{tabular}
    \caption{{\bf Epidemic models.} Average (variance in parenthesis) of the 1,000 parameter estimates based on data generated from the deterministic and the stochastic SIR models \eqref{model SIR}. True values are $\alpha = 0.5$, $ \beta = 0.3$.}
    \label{table SIR no perturbation}
\end{table}

Data sets were simulated from the deterministic and stochastic SIR models \eqref{model SIR}, with parameters $\alpha = 0.5$, $ \beta = 0.3$. $ \sigma_1 = 3\times10^{-3}$, $ \sigma_2 = 1\times10^{-3}$, $\gamma_1 = \sigma_1\sqrt{T}$, $ \gamma_2 = \sigma_2/\sqrt{2(\beta - \alpha s^*)}$ and initial values $s_0 = 0.99, i_0 = 0.001$. The long-term equilibrium is denoted $s^*$, which is calculated from $\frac{\alpha}{\beta}(1-s^*) + \ln\frac{s^*}{s_0} = 0$,  and $i^* = 0$. We restarted all simulations that ended prematurely (i.e., the number of infections reached zero before the end of the simulation), resulting in 1,000 datasets for both models. We fitted ODE and SDE models to each dataset. Density plots of the estimates of $\alpha$ and $\beta$ are shown in Figure \ref{fig: SIR original data set} and the average and variance of estimates can be found in Table \ref{table SIR no perturbation}.

For data generated from the deterministic SIR model, both ODE and SDE estimates seem unbiased, however, the SDE model has larger variance. For data generated from the stochastic SIR model, the SDE fit has a negligible bias, whereas the ODE fit has a bias of around 20\% in both parameters.

\paragraph{SIR model with instantaneous perturbation.}
We added an instantaneous perturbation $h = (-0.02,0.02)^\top$ at time $t_p = 10$ and simulated 1,000 data sets from the deterministic and the stochastic SIR model \eqref{model SIR}, respectively. We compare to data without perturbations by using the same set of random numbers in the simulations. Parameters are set to $\alpha = 0.5$, $ \beta = 0.3$, $ \sigma_1 = 3\times10^{-3}$, $ \sigma_2 = 1\times10^{-3}$, $\gamma_1 = \sigma_1\sqrt{T}$, $ \gamma_2 = \sigma_2/\sqrt{2(\beta - \alpha s^*)}$  and initial values are $s_0 = 0.99, i_0 = 0.001$.

\begin{figure}[!t]
    \centering
    \includegraphics[width = 0.95\textwidth]{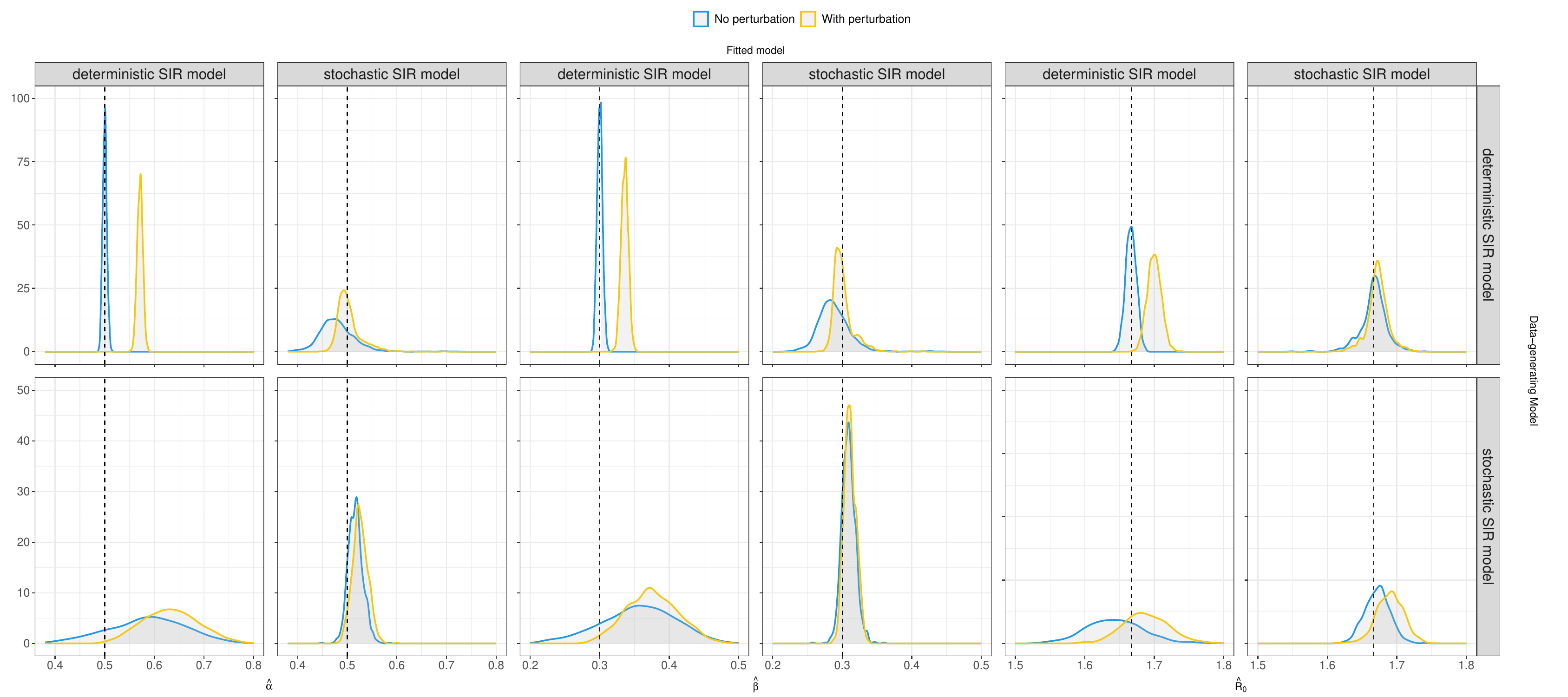}
    \caption{{\bf SIR model with instantaneous perturbation}. Density plots of the estimates from 1,000 data sets of the contact rate $\alpha$ and the removal rate $\beta$ in the deterministic or the stochastic SIR model \eqref{model SIR}. Parameters are $\alpha = 0.5$, $ \beta = 0.3$, $ \sigma_1 = 3\times10^{-3}$, $ \sigma_2 = 1\times10^{-3}$,  $\gamma_1 = \sigma_1\sqrt{T}$, $ \gamma_2 = \sigma_2/\sqrt{2(\beta - \alpha s^*)}$, initial values are $s_0 = 0.99, i_0 = 0.001$. Dashed lines are at the true values used in the simulation.
    }
    \label{fig: SIR ptb sim}
\end{figure}

Figure \ref{fig: SIR ptb sim} shows density plots of the estimates with different data-generating models and models used for fitting. The perturbations affect parameter estimates in all scenarios. Using an ODE model always introduces a larger bias, regardless of the data-generating model. Using the correct model decreases the estimator variance.

\paragraph{SEIR model.}
\begin{figure}[ht]
    \centering
    \includegraphics[width = 0.8\textwidth]{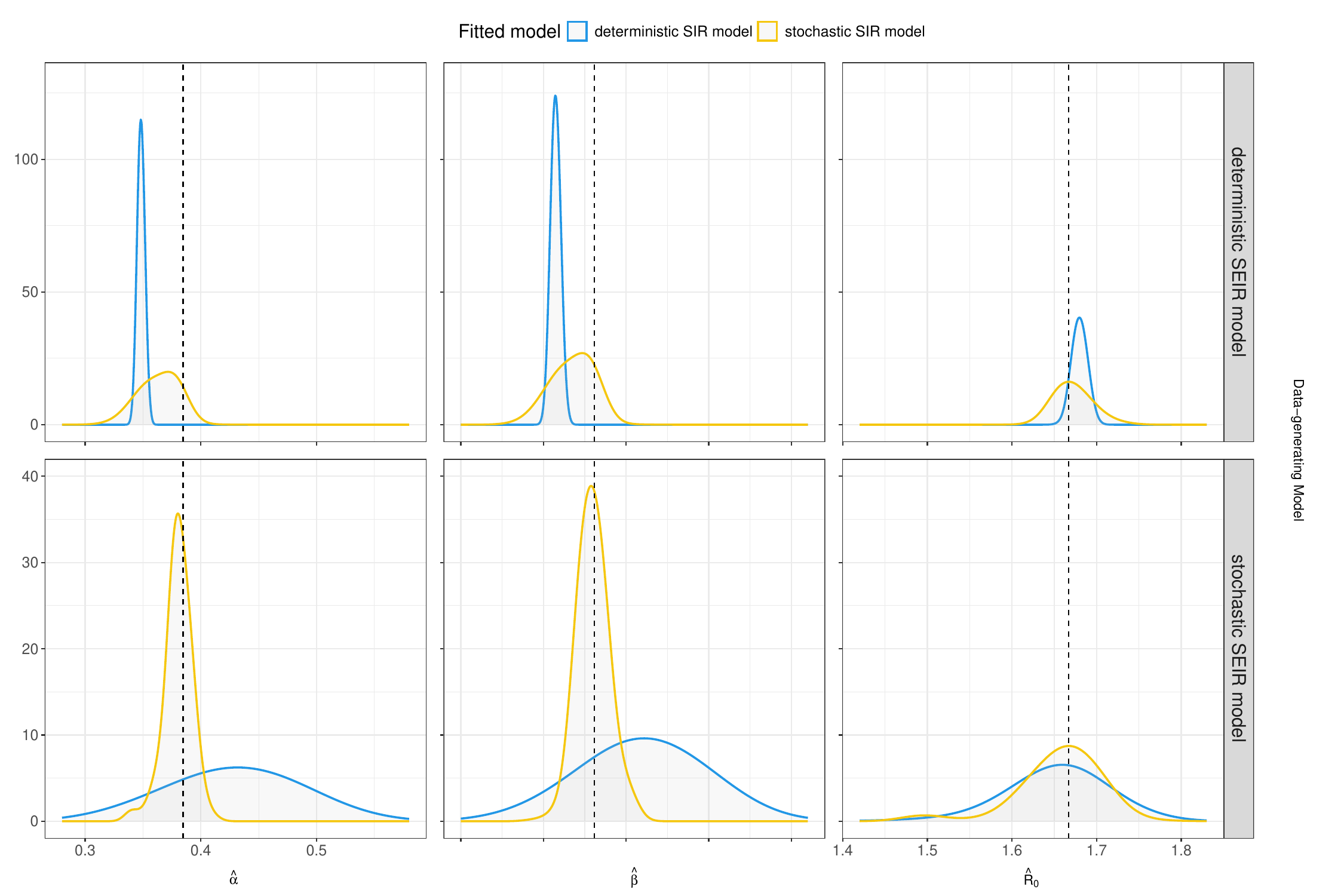}
    \caption{{\bf SEIR model.} Density plots based on 1,000 estimates of the contact rate $\alpha$, the removal rate $\beta$ and the basic reproduction number ${R}_0 = {\alpha}/{\beta}$ from data sets simulated from the deterministic and the stochastic SEIR model \eqref{SEIR model} with $\Delta = 0.5$ and $T = 40$. Parameters are $\alpha = 0.5$, $ \lambda = 1$, $\beta = 0.3$, $ \sigma_1 = 3\times10^{-3}$, $ \sigma_2 =\sigma_3 = 1\times10^{-3}$, $ \gamma_1 = \sigma_1\sqrt{T}$, $ \gamma_2 =\frac{\sigma_2}{\sqrt{2\lambda}}$, and $\gamma_3 =\frac{\sigma_3}{\sqrt{2\beta}}$, initial values are $s_0 = 0.99, e_0 = i_0 = 0.001$. Dashed lines are $x = \frac{\lambda}{\lambda + \beta} \alpha$ and $x = \frac{\lambda}{\lambda + \beta} \beta$.}
    \label{fig: SEIR sim}
\end{figure}

We simulated 1,000 data sets from both the deterministic and stochastic SEIR models \eqref{SEIR model}, using $\Delta = 0.5$ and $T = 40$.
Parameters are $\alpha = 0.5$, $ \lambda = 1$, $\beta = 0.3$, $ \sigma_1 = 3\times10^{-3}$, $ \sigma_2 =\sigma_3 = 1\times10^{-3}$, $ \gamma_1 = \sigma_1\sqrt{T}$, $\gamma_2 =\frac{\sigma_2}{\sqrt{2\lambda}} $, and $\gamma_3 = \frac{\sigma_3}{\sqrt{2\beta}} $ and initial values are $s_0 = 0.99, e_0 = i_0 = 0.001$. The parameters have the relationship $\alpha' = \frac{\lambda}{\lambda + \beta} \alpha$ and $\beta' = \frac{\lambda}{\lambda + \beta} \beta$ (Section\ref{sec:SEIR model}). In Figure \ref{fig: SEIR sim} we therefore use $\alpha'$ and $\beta'$ as baselines. We fitted the simpler SIR model to each data set and estimated the contact rate $\alpha'$ and the removal rate $\beta'$. 

Figure \ref{fig: SEIR sim} shows density plots of the 1,000 estimates of $\alpha$ and $\beta$ for each estimator. Using the correct model (ODE or SDE) leads to smaller estimator variance, whereas estimates are biased due to the model misspecification.
However, this bias is negligible when using the SDE model for fitting, and thus illustrates how a stochastic model can compensate for model misspecifications. 

In epidemic modelling, a key quantity of interest is the basic reproduction number $R_0$. In the SEIR model \eqref{SEIR model}, it is given by $R_0 = \alpha' / \beta' = 5/3$. The densities of the estimates of $R_0$ are plotted in Fig. \ref{fig: SEIR sim}, 
where we clearly see that no matter if the data-generating model is deterministic or stochastic, the use of the stochastic SIR model for fitting always provides less biased estimates of the true basic reproduction number $R_0$ compared to the deterministic model.

\section{COVID-19 in Denmark}\label{application}
\begin{figure}[ht]
    \centering
    \includegraphics[width = 0.8\textwidth]{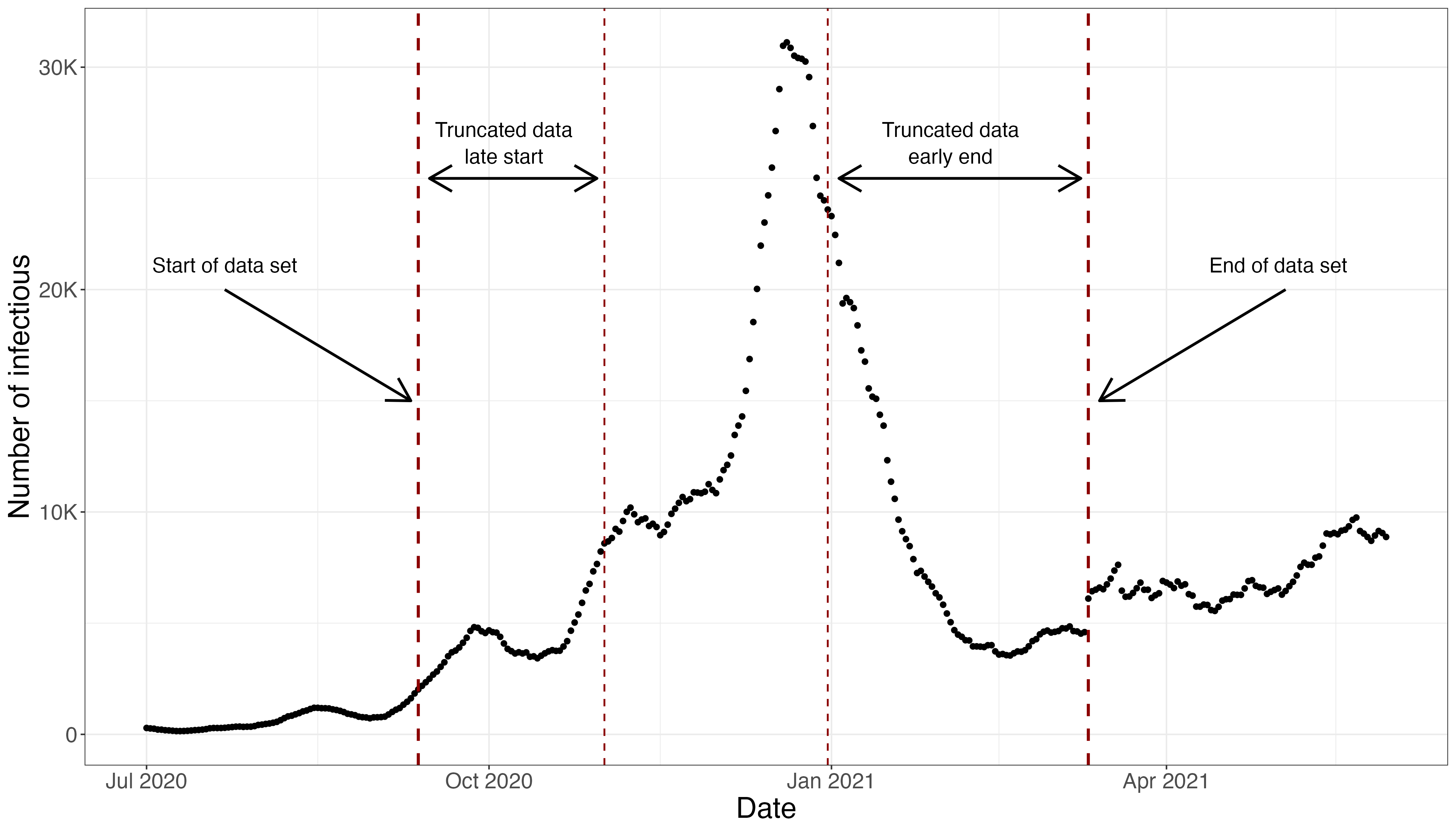}
    \caption{{\bf Number of COVID-19 infectious cases from 2020-05-10 to 2021-08-03 in Denmark.} Data points are number of people that tested positive cumulated over the last nine days, providing an estimate of number of infectious at any given day, assuming that an infected person stays infectious for nine days. Observations between the outer dashed lines are the full data of the third wave used in the fitting, from 2020-09-12 to 2021-03-11. The inner dashed lines indicate the limits for sequentially removing observations from either the beginning or the end of the record. The truncated data sets were used for checking robustness of estimates.}
    \label{fig: covid 19 denmark}
\end{figure}

The first case of infection in Denmark during the COVID-19 pandemic was reported on February 17, 2020. Here, we analyze data from the third wave in Denmark (September 12, 2020 to March 11, 2021), illustrated in Fig. \ref{fig: covid 19 denmark} and fit both deterministic and stochastic SIR models (Section \ref{Example 2}). To fit the models, we use the accumulated number of infections reported in the 9 days prior and up to the current day as the number of infectious, assuming that an infected person stays infectious for 9 days \cite{TO2020565}. We assumed a Danish population of 5.86 million \cite{denmarkpopulation} to calculate the proportion of daily infections.

\subsection{Parameter estimates}
\label{corona denmark parameter estimates}
Suppose that all infections are measured, i.e., there is no dark figure. This is of course not realistic, however, Denmark had a very extensive testing policy in that period of the pandemic, and few cases have been missed. The main sources of measurement error are false negatives and false positives. We use the proportion of infected cases on September 12, 2020 as initial condition $i_0$. The cumulative proportion of infected cases from September 3, 2020 to March 2, 2021, i.e., 9 days before the start date, is used as initial condition $r_0$. The initial value for the proportion of susceptibles is therefore $s_0 = 1 - i_0 - r_0$. 

To mitigate the effects of local minima, we used the \texttt{gosolnp} global optimisation procedure with 500 random initialisations in fitting both deterministic and stochastic models. Observations are partial, since only $i_t$ is observed. To fit the deterministic SIR model, we therefore did least squares only on the $I$ compartment.  The estimates were $\hat{\alpha}_{ode} = 0.697$ and $\hat{\beta}_{ode} = 0.633$ and the reproduction number was thus $\hat R_{0}^{ode} = \hat{\alpha}_{ode} / \hat{\beta}_{ode} = 1.101$.

For the stochastic SIR model, we could not use the estimation method \eqref{Strang splitting scheme} because only partial observations were available. Instead we used the Unscented Kalman Filter (UKF) to approximate the likelihood function.
We extended the SDE model to include measurement error to avoid the updated state covariance being singular in the UKF, assuming normally distributed measurement errors. 
The parameter estimates corresponding to the minimum achieved objective value were $\hat{\alpha}_{sde} = 0.736$ and $\hat{\beta}_{sde} = 0.640$.
The reproduction number was $\hat R_{0}^{sde} = \hat{\alpha}_{sde} / \hat{\beta}_{sde} = 1.151$. The ODE and the SDE models thus approximately agree on parameter estimates though the $R_0$ is larger in the SDE model.

\subsection{Stability of parameter estimates}\label{stability SIR}
\begin{figure}[!t]
    \centering
    \includegraphics[width = 0.8\linewidth]{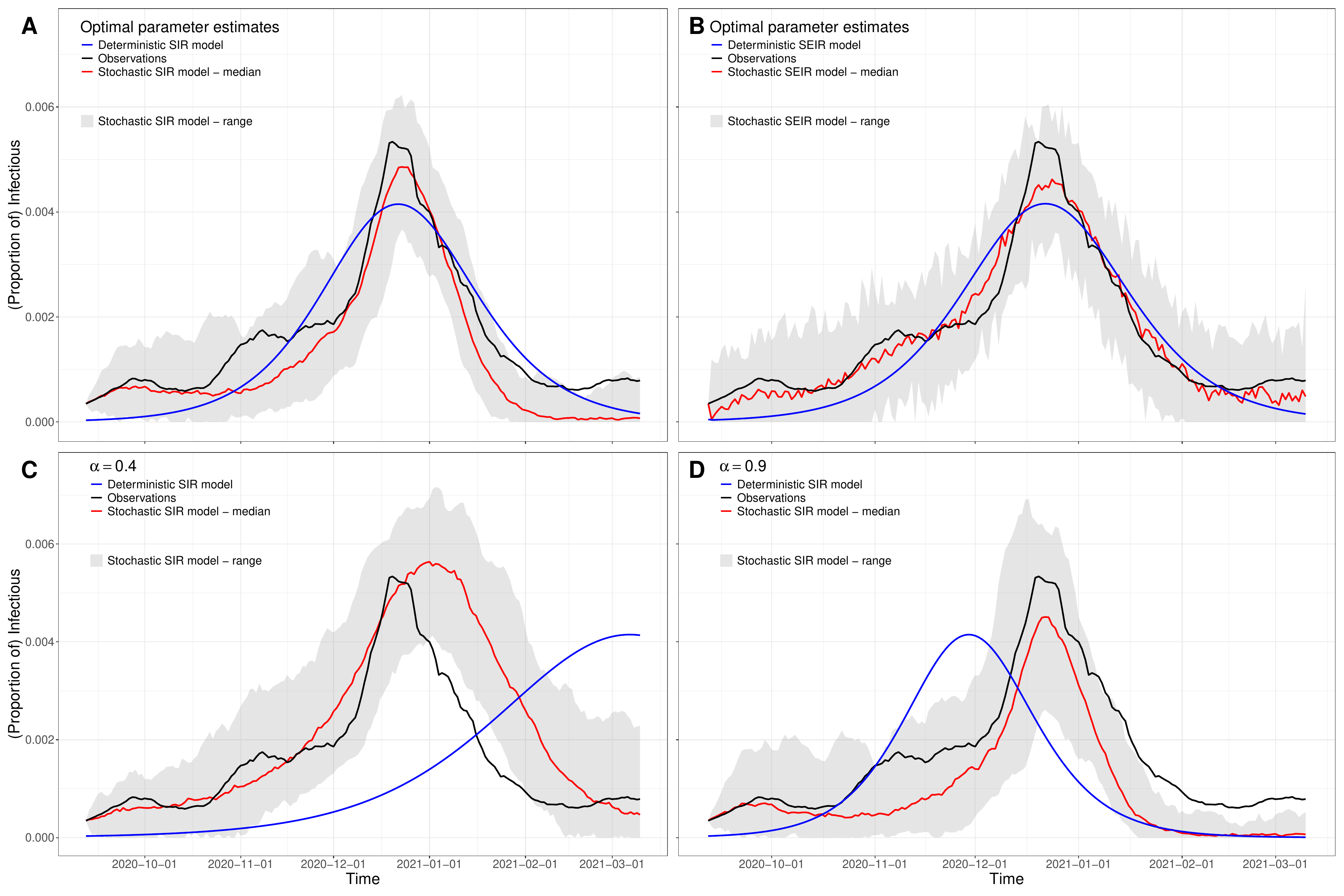}
    \caption{{\bf Predictions of the proportion of infectious.} The deterministic model (dark blue), the median of the stochastic model (dark red) and the range of 50 simulations (gray area) with different parameter values, and the real data (black). A, C and D: SIR model, generated using the same set of random seeds. B: SEIR model. A and B: The parameter values used are the optimal estimates. A: $\hat{\alpha}_{ode} = 0.697$, $\hat{\beta}_{ode} = 0.633$, $\hat{\alpha}_{sde} = 0.736$, $\hat{\beta}_{sde} = 0.640$. B: $\hat{\alpha}_{ode} = 0.998$, $\hat{\gamma}_{ode}= 0.895$, $\hat{\beta}_{ode} = 0.871$, $\hat{\alpha}_{sde} = 0.973$, $\hat{\gamma}_{sde}= 0.501$, $\hat{\beta}_{sde} = 0.839$. C: $\alpha_{ode} = {\alpha}_{sde} = 0.4$. D: $\alpha_{ode} = \alpha_{sde} = 0.9$. C and D: ${\alpha}_{ode} / {\beta}_{ode} = 1.101$ and ${\alpha}_{sde} / {\beta}_{sde} = 1.151$. }
    \label{fig: prediction uptdate}
\end{figure}

Figure \ref{fig: prediction uptdate} A and B show predictions from the fitted ODE and SDE models. 
We also set the parameter values to values differing from the optimal values while maintaining the reproduction number at its optimal value for each model, results are in Fig. \ref{fig: prediction uptdate} C and D. When the parameter values are far from the optimum, the ODE model is even further from the data, whereas the UKF of the SDE model still gives good predictions, illustrating the resilience of the SDE model to model misspecifications.

Furthermore, we investigated the impact of changing the length of the observation interval on the robustness of the ODE and the SDE estimators by changing either the start date or the end date of the data interval. First, we sequentially deleted the last data point until the last 70 data points were removed, thus obtaining 70 data sets of differing sizes. The end date of the smallest data set was December 31, 2020, see Fig. \ref{fig: covid 19 denmark}. On each data set, parameter estimates were obtained for both models, illustrating how estimates would have been at that point in the epidemic. Likewise, we deleted sequentially one observation from the beginning up to 50 observations, so that the data set starts from October 31, 2020. This illustrates the situation where the surveillance of the epidemic starts later.

\begin{figure}[!t]
    \centering
    \includegraphics[width = 0.9\textwidth]{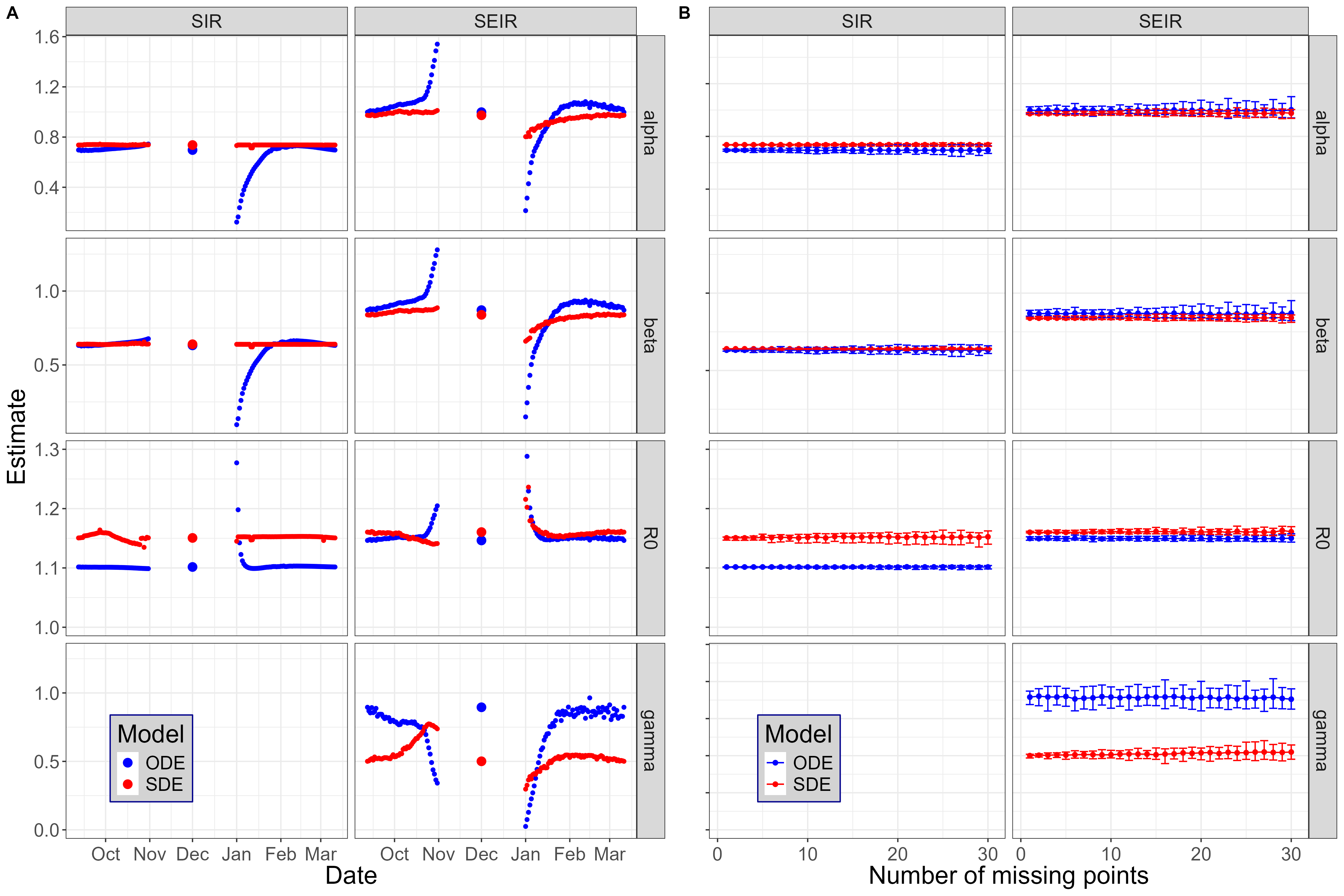}
    \caption{{\bf Comparison of parameter estimates using different models and reduced data sets of number of Covid-19 infectious in Denmark.} Four models were used for fitting the data: deterministic and stochastic SIR and SEIR models. The full data set are number of people tested positive in the last 9 nine days, between 12 September 2020 and 11 March 2021. A. Either the last points or the first points in the dataset were removed sequentially, up to 70 data points were removed in the end or 50 data points were removed in the beginning of the data set. The larger dots are the estimates from the full data set. Estimates using the deterministic models are sensitive to the length of the data set as well as to the chosen epidemic model, the stochastic models are more robust. B. The average and range of parameter estimates over 100 (SIR models) or 50 (SEIR models) data sets where $k$ observations, $k=1,...,30$, were removed randomly from the dataset. The variance of the estimates increases more for deterministic compared to stochastic models. 
    }
    \label{fig: delete points}
\end{figure}

For each truncated data set, the optimal values of the deterministic SIR model were used as starting values for the UKF to fit the stochastic SIR model, results are presented in Fig. \ref{fig: delete points}. Clearly, parameter estimates from the ODE model are much more sensitive to the length of the data set, no matter if it is the beginning or the end of the wave that are missing, making estimates less trustworthy, whereas the SDE model seems to give robust parameter estimates (in the sense of not changing much). 

Finally, we consider the effect of randomly deleting observation points on the parameter estimates. We randomly removed $k$ observations from the complete dataset, for $k = 1, \ldots , 30$. For each $k$, we sampled new incomplete data sets 100 times, thus obtaining a total of 3,000 incomplete data sets. On each incomplete data set, we estimated parameters in both models. Fig. \ref{fig: delete points} illustrates the average and the range of the estimates for each $k$. The range tends to increase for both the deterministic and stochastic SIR models as $k$ increases, as expected, whereas the average is stable. However, the range of the SDE estimates are smaller than the ODE estimates of $\alpha$ and $\beta$ for all $k$, once again highlighting the robustness of using an SDE model for estimation.

\subsection{The SEIR model}

We extended the SIR model to an SEIR model to explore whether the same effects are present. We used the same data sets and assumed that the initial exposed cases were half of the infected cases. Thus, $e_0 = 0.5i_0$ and $s_0 = 1 - e_0 - i_0 - r_0$.

Parameters are estimated similarly as for the SIR model. In the deterministic SEIR model, the estimates are $\hat{\alpha}_{ode} = 0.998$, $\hat{\gamma}_{ode}= 0.895$, and $\hat{\beta}_{ode} = 0.871$, and thus, $\hat R_0^{ode} = 1.146$. In the stochastic SEIR model, the estimates are $\hat{\alpha}_{sde} = 0.973$, $\hat{\gamma}_{sde}= 0.501$ and $\hat{\beta}_{sde} = 0.839$, and thus, $\hat R_0^{sde} = 1.157$. 
Both the ODE and SDE versions of the SEIR model have an increased basic reproduction number, and only the SDE version of the SEIR model provides an estimate of the basic reproduction number that is close to the estimates provided by the SIR model. Moreover, the estimate from the SDE model is more in agreement with the official estimates in Denmark at that point in time of the pandemic (it was 1.16 in January 2021, see \cite{SSIrapport} (in Danish)). 

We repeated the estimation on the incomplete data sets. We observe similar results as in the SIR model, results are shown in Fig. \ref{fig: delete points}A. In the deterministic SEIR model, estimates of $\alpha$,  $\gamma$ and $\beta$ change dramatically when observations are removed, as in the SIR model. The stochastic SEIR model still provides robust estimates, except for estimates of $\gamma$. 

Finally, we consider the effect of randomly deleting observation points on the parameter estimates as in Section \ref{stability SIR}, however, only using the first 50 sampled data sets for each $k$ due to the SEIR model taking much longer time to run than the SIR model. Results are in Fig. \ref{fig: delete points}B, where clearly the estimates from the deterministic SEIR model are more unstable with increasing amount of missing data, whereas the estimates from the stochastic model are stable. 

\section{Conclusion}\label{Conclusion}
We investigated the performance of ODE and SDE models with additive noise in terms of accuracy and robustness of parameter estimation. In simulation studies as well as in real-life applications, we show the significant advantage of using an SDE model over an ODE model if there is any misspecification of the model.

It is known that using a stochastic model for parameter estimation in highly nonlinear models regularizes the likelihood and thus improve parameter estimates, even if the data-generating mechanism is an ODE model \cite{Leander:2014}. This is because the likelihood of the stochastic model usually has less optima than the likelihood from the deterministic model. In linear models, this is not an issue. However, here we show that even for linear models, the stochastic models improve the estimation problem if the data are subject to some perturbations not captured by the model.

By analysing simple one-dimensional linear systems as well as more complex epidemic models (deterministic and stochastic SIR and SEIR models), we systematically evaluated the impact of assuming either a deterministic or a stochastic model on the statistical inference. We explored a variety of model misspecification scenarios that allow data to be generated from more complex data-generating mechanisms than the models used for fitting. Model misspecifications cause estimation errors, bias, and increased variance in the fitting of both the ODE and SDE models. However, we found that fitting the SDE model provided more stable and less biased parameter estimates. This advantage is even more noticeable when systematic errors are present in the data. The application to real data from COVID-19 in Denmark further supports these conclusions. 

Another key factor that we investigated was the effect of the length of the time series on parameter estimates. By fitting deterministic and stochastic SIR and SEIR models, we found that the SDE model estimates remain stable under missing data scenarios, while the ODE model is more sensitive to individual data points. Even for short time series in the COVID-19 data only covering either the beginning or the end of an infection wave, the SDE model provides reliable parameter estimates, whereas the ODE model is highly variable. This is important for practical applications where data availability is often limited. This reinforces the idea that the inclusion of stochasticity can improve the robustness of statistical inference in epidemic modelling and other fields, especially when only incomplete data are available.

In summary, we emphasize the need for careful consideration of model choice in parameter estimation. Although ODE models are useful in some situations, they may not be sufficient when dealing with noisy or complex systems. Our findings suggest that SDE models provide a more reliable framework for capturing real-world variability, reducing estimation bias, and improving inference under uncertainty.

\section{Technical results}\label{Calculation}
\subsection{Taylor expansion of a ratio of random variables}
\label{Taylor expansion}
Let $X_1,X_2$ be two random variables. A Taylor expansion up to second order of $X_1/X_2$ around $\mathbb{E}(X_1)$ and $\mathbb{E}(X_2)$ and taking expectations yield

\begin{align}
&\begin{aligned}
   \label{mean x1/x2}
   \mathbb{E}\left (\frac{X_1}{X_2} \right )  = \frac{\mathbb{E}(X_1)}{\mathbb{E}(X_2)} -
   \frac{\mbox{Cov}(X_1,X_2)}{\mathbb{E}(X_2)^2} +
   \frac{\mathbb{E}(X_1)V(X_2)}{\mathbb{E}(X_2)^3}
\end{aligned}  
\intertext{and}
&\begin{aligned}
   \label{variance x1/x2}
   V\left(\frac{X_1}{X_2}\right) =  \mathbb{E}\left(\frac{X_1^2}{X_2^2}\right) -  \mathbb{E}\left(\frac{X_1}{X_2}\right)^2 \approx \frac{V(X_1)}{\mathbb{E}(X_2)^2}  -\frac{2\mathbb{E}(X_1)\mbox{Cov}(X_1,X_2)}{\mathbb{E}(X_2)^3}
   +\frac{\mathbb{E}(X_1)^2V(X_2)}{\mathbb{E}(X_2)^4}.
\end{aligned}
\end{align}

\subsection{Moments} \label{sec:moments}
The expectation and variance of $Y^{sde}$ and $Z^{sde}$ in \eqref{Y & Z sde} are given in \cite{ditlevsen:2020} as 
\begin{align}
\label{E_Y sde}
& \begin{aligned}
\mathbb{E}(Y^{sde}) &= \phi(n) + \left(
n -  \frac{1-\rho^{2n}}{1-\rho^2}
\right)
\frac{\sigma^2}{2a},\\
\end{aligned}
\\[1ex]
\label{E_Z sde}
& \begin{aligned}
\mathbb{E}(Z^{sde}) &= \rho \phi(n) + 
\rho \left(
n -  \frac{1-\rho^{2n}}{1-\rho^2}
\right)\frac{\sigma^2}{2a},\\
\end{aligned}
\\[1ex]
\label{V_Y sde}
& \begin{aligned}
V(Y^{sde}) = & 
\frac{\sigma^2}{2a}\left(
4\phi(n) \frac{\rho^2+\rho^{2n}}{1-\rho^2}
- 8n\frac{(x_0-b)^2\rho^{2n}}{1-\rho^2}
\right)+\\
&\frac{\sigma^4}{4a^2}
\left(
\frac{2(1+\rho^2)+8\rho^{2n}}{1-\rho^2}n-
\frac{2(1-\rho^{2n})(1+2\rho^2+\rho^{2n})}{(1-\rho^2)^2}
\right),
\end{aligned}
\\[1ex]
\label{V_Z sde}
& \begin{aligned}
V(Z^{sde})= & \frac{\sigma^2}{2a}\phi(n)
\left(
\frac{(1+\rho^2)^2-4\rho^{4n+2}}{(1-\rho^2)(1-\rho^{2n})}
-\frac{(1-\rho^2)\rho^{2n}+4n\rho^{2n}(1+\rho^2)}{1-\rho^{2n}}
\right) +\\
&  \frac{\sigma^4}{4a^2} 
\left(
\frac{1+4\rho^2-\rho^4+4\rho^{2n}(1+\rho^2)}{1-\rho^2}n
-\frac{(1+\rho^2)^2+6\rho^2-\rho^{2n}(1-\rho^2)^2-2\rho^{4n+2}}{(1-\rho^2)^2}
\right),
\end{aligned}
\\[1ex]
\label{Cov_Y_Z sde}
& \begin{aligned}
\mbox{Cov}(Y^{sde},Z^{sde}) =& 2\frac{\sigma^2}{2a}\phi(n)\rho\left(
\frac{1+\rho^2-2\rho^{4n}}{(1-\rho^2)(1-\rho^{2n})}-
\frac{(1+3\rho^2)n\rho^{2n-2}-\rho^{2n}}{1-\rho^{2n}}
\right)+\\
&\frac{\sigma^4}{4a^2} 2\rho\left(
\frac{2+(1+3\rho^2)\rho^{2n-2}}{1-\rho^2}n-\frac{1-2\rho^{2n}-\rho^{4n}}{(1-\rho^2)^2}
\right).
\end{aligned}
\end{align}

Define $Y^{ode} = \sum_{i=1}^n (y_{i-1}^{ode}-b)^2$ and $Z^{ode} = \sum_{i=1}^n (y_i^{ode}-b)(y_{i-1}^{ode}-b)$. Using \eqref{example 1 ODE}, we obtain
\begin{align} 
\label{E_Y ode}
& \begin{aligned}
\mathbb{E}(Y^{ode}) &= \phi(n)+n\sigma_0^2,\\
\end{aligned}
\\[1ex]
\label{E_Z ode}
& \begin{aligned}
\mathbb{E}(Z^{ode}) &= \rho\phi(n),\\
\end{aligned}
\\[1ex]
\label{V_Y ode}
& \begin{aligned}
V(Y^{ode}) &= 4\sigma_0^2 \phi(n)+2n\sigma_0^4,
\end{aligned}
\\[1ex]
\label{V_Z ode}
& \begin{aligned}
V(Z^{ode}) &= \sigma_0^2\phi(n)\frac{1+3\rho^2- 3\rho^{2n} -\rho^{2n+2}}{1-\rho^{2n}}
+n\sigma_0^4,\\
\end{aligned}
\\[1ex]
\label{Cov_Y_Z ode}
& \begin{aligned}
\mbox{Cov}(Y^{ode},Z^{ode})
&= 2\sigma_0^2 \phi(n)\frac{2\rho-\rho^{2n-1}-\rho^{2n+1}}{1-\rho^{2n}}.
\end{aligned}
\end{align}

{\bf Moments of $Y^{ode}_1$ and $Z^{ode}_1$ and approximation of $\mathbb{E}(\hat{\rho}_1^{ode})$}. \label{Calculation ode instantaneous}
Define $Y^{ode}_1 = \sum_{i=1}^n (y_{i-1}^{ode}-b)^2, \quad  Z^{ode}_1 = \sum_{i=1}^n (y_i^{ode}-b)(y_{i-1}^{ode}-b)$ and observations $\{y^{ode}_0,y^{ode}_1,...,y^{ode}_n\}$ from the ODE model \eqref{model instantaneous} with instantaneous perturbation. The expectation and variance of $Y^{ode}_1$ and $Z^{ode}_1$ are for $k < n$ 
\begin{align}
    &\begin{aligned}
    \mathbb{E}(Y^{ode}_1) & =  \mathbb{E}(Y^{ode}) 
    + \frac{\rho^{2k}-\rho^{2n}}{1-\rho^2}
    \left(
    2(x_0 -b)e^{at_p}h + e^{2at_p}h^2
    \right),
    \end{aligned}
    \\[1ex]
    &\begin{aligned}
    \mathbb{E}(Z^{ode}_1) &= \mathbb{E}(Z^{ode}) +
    \frac{\rho^{2k}-\rho^{2n}}{1-\rho^2}\left(
    \frac{\rho^{2k-1}(1+\rho^2)-2\rho^{2n+1}}{\rho^{2k}-\rho^{2n}}(x_0-b)e^{at_p}h
    +\rho e^{2at_p}h^2
    \right),
    \end{aligned}
    \\[1ex]
    &\begin{aligned}
    V(Y^{ode}_1) & = V(Y^{ode}) 
    + 4 \sigma_0^2 \frac{\rho^{2k}-\rho^{2n}}{1-\rho^2}
    \left(
    2 (x_0 -b)e^{at_p}h + e^{2at_p}h^2
    \right),
    \end{aligned}
    \\[1ex]
    &\begin{aligned}
    V(Z^{ode}_1)  = & V(Z^{ode}) 
    + \\
    &\sigma_0^2 \left(
    \frac{\rho^{2k-2}(1+\rho^2)^2 - 3\rho^{2n} -\rho^{2n+2}}{1-\rho^2}
    2(x_0 -b)e^{at_p}h +\frac{2\rho^{2k}(1+\rho^2) - 3\rho^{2n} -\rho^{2n+2}}{1-\rho^2}
    e^{2at_p}h^2
    \right),
    \end{aligned}
    \\[1ex]
    &\begin{aligned}
    \mbox{Cov}(Y^{ode}_1 ,Z^{ode}_1) 
    = & \mbox{Cov}(Y^{ode},Z^{ode}) + \\
    &  2\sigma_0^2 \left(
    \frac{2(1+\rho^2)(\rho^{2k}-\rho^{2n})}{\rho(1-\rho^2)}(x_0-b) e^{at_p}h 
    + \frac{4\rho^{2k+2}-2(1+\rho^2)\rho^{2n}}{\rho(1-\rho^2)}e^{2at_p}h^2
    \right).
    \end{aligned}
\end{align}
Moments of $Y^{ode}$ and $Z^{ode}$ are given in \eqref{E_Y ode}--\eqref{Cov_Y_Z ode}. For a perturbation at $ t_p \in (t_{n-1}, t_n)$ (only the last observation is affected), the moments are
\begin{align*}
    & \mathbb{E}(Y^{ode}_1) = \mathbb{E}(Y^{ode}),\\
    & \mathbb{E}(Z^{ode}_1) = \mathbb{E}(Z^{ode}) + (x_0-b)e^{at_p}\rho^{2n-1}h,\\
    & V(Y^{ode}_1)  =  V(Y^{ode}), \\
    & V(Z^{ode}_1)  =  V(Z^{ode}) + 2 \sigma_0^2 (x_0-b)\rho^{2n-2}(1+\rho^2)e^{at_p}h + \sigma_0^2\rho^{2n} e^{2at_p} h^2,\\
    & \mbox{Cov}(Y^{ode}_1 ,Z^{ode}_1) = \mbox{Cov}(Y^{ode} ,Z^{ode})+ 2 \sigma_0^2 (x_0-b)\rho^{2n-1} e^{at_p} h.
\end{align*}
Using the moments of $Y^{ode}_1$ and $Z^{ode}_1$, we approximate the expectation of $\hat{\rho}^{ode}_1$ with a Taylor expansion up to second order of $h$ around $h=0$. Then $A_1(\rho, n,k, \sigma_0^2)$ and $A_2(\rho, n,k, \sigma_0^2)$ in \eqref{E rho_1 ode approx} have the form when $k < n$
\begin{align*}
    &\begin{aligned}
     A_1(\rho, n,k, \sigma_0^2) =  & \frac{(x_0-b)e^{at_p}}{\mathbb{E}(Y^{ode})}\frac{\rho^{2k}-\rho^{2n}}{1-\rho^2}
     \left(
     \frac{\rho^{2k-1}(1+\rho^2)-2\rho^{2n+1}}{\rho^{2k}-\rho^{2n}}
     -\frac{2\mathbb{E}(Z^{ode})+\frac{4(1+\rho^2)}{\rho}\sigma_0^2}{\mathbb{E}(Y^{ode})}+
     \right. \\
     & \left.
     \frac{4\mbox{Cov}[Y^{ode} ,Z^{ode}) + \frac{\rho^{2k-1}(1+\rho^2)-2\rho^{2n+1}}{\rho^{2k}-\rho^{2n}} V(Y^{ode}) + 8\sigma_0^2 \mathbb{E}(Z^{ode}) }{(\mathbb{E}(Y^{ode}))^2}
     -\frac{6\mathbb{E}(Z^{ode}) V(Y^{ode})}{(\mathbb{E}(Y^{ode}))^3}
     \right),
    \end{aligned}
    \\[1ex]
    &\begin{aligned}
    A_2(\rho, n,k, \sigma_0^2) =   & \frac{e^{2at_p}}{\mathbb{E}(Y^{ode})} 
    \frac{\rho^{2k}-\rho^{2n}}{1-\rho^2}
    \left(
    \rho - \frac{\mathbb{E}(Z^{ode})+4\sigma_0^2\frac{2\rho^{2k+2}-(1+\rho^2)\rho^{2n}}{\rho(\rho^{2k}-\rho^{2n})}}{\mathbb{E}(Y^{ode})}
    + \right. \\
    & \left.\frac{2\mbox{Cov}[Y^{ode} ,Z^{ode})+\rho V(Y^{ode})+4\sigma_0^2\mathbb{E}(Z^{ode})}{(\mathbb{E}(Y^{ode}))^2}
    -\frac{3\mathbb{E}(Z^{ode})V(Y^{ode})}{(\mathbb{E}(Y^{ode}))^3}
    \right) -
    \\
     & 2e^{2at_p}
    \left(
    \frac{\rho^{2k-1}(1+\rho^2)-2\rho^{2n+1}}{\rho^{2k}-\rho^{2n}}
    - 
    \frac{2\mathbb{E}(Z^{ode}) + \left(8\frac{1+\rho^2}{\rho}+\frac{4\rho^{2k-1}(1+\rho^2)-8\rho^{2n+1}}{\rho^{2k}-\rho^{2n}}\right)\sigma_0^2}
    {\mathbb{E}(Y^{ode})}+
    \right.\\
    &\frac{6\mbox{Cov}[Y^{ode} ,Z^{ode})+\frac{3\rho^{2k-1}(1+\rho^2)-6\rho^{2n+1}}{\rho^{2k}-\rho^{2n}}V(Y^{ode}) + 24\sigma_0^2}{(\mathbb{E}(Y^{ode}))^2}
    - \\
    &\left.\frac{12\mathbb{E}(Z^{ode})V(Y^{ode})}{(\mathbb{E}(Y^{ode}))^3}
    \right)
   \frac{ (x_0-b)^2}{(\mathbb{E}(Y^{ode}))^2}
    \left(\frac{\rho^{2k}-\rho^{2n}}{1-\rho^2}\right)^2.
    \end{aligned}
\end{align*}
For $a\Delta \ll 1$,  $A_1(\rho, n,k, \sigma_0^2)$ and $A_2(\rho, n,k, \sigma_0^2) > 0 $ for large $n$ and $n-k$.

Define $Y^{sde}_1 = \sum_{i=1}^n (y_{i-1}^{sde}-b)^2, \quad  Z^{sde}_1 = \sum_{i=1}^n (y_i^{sde}-b)(y_{i-1}^{sde}-b)$ and assume observations $\mathbf{y}^{sde}$ from the ODE model \eqref{model instantaneous} with instantaneous perturbation. The expectation and variance of $Y^{sde}_1$ and $Z^{sde}_1$ for $k < n$  are
\begin{align}
    \label{EY sde 1}
    &\begin{aligned}
    \mathbb{E}(Y^{sde}_1)  =  &\mathbb{E}(Y^{sde}) 
    + \frac{\rho^{2k}-\rho^{2n}}{1-\rho^2}
    \left(
    2(x_0 -b)e^{at_p}h + e^{2at_p}h^2
    \right),
    \end{aligned}
    \\[1ex]
    \label{EZ sde 1}
    &\begin{aligned}
    \mathbb{E}(Z^{sde}_1) = &\mathbb{E}(Z^{sde}) +
    \frac{\rho^{2k}-\rho^{2n}}{1-\rho^2}\left(
    \frac{\rho^{2k-1}(1+\rho^2)-2\rho^{2n+1}}{\rho^{2k}-\rho^{2n}}(x_0-b)e^{at_p}h
    +\rho e^{2at_p}h^2
    \right),
    \end{aligned}
    \\[1ex]
    \label{VY sde 1}
    &\begin{aligned}
    V(Y^{sde}_1)  = &V(Y^{sde}) 
    +8(x_0 -b) e^{at_p}\frac{\sigma^2}{2a}
    \left(
    \frac{\rho^{2k}-(2n-2k+1)\rho^{2n}(1-\rho^2)-\rho^{4n-2k+2}}{(1-\rho^2)^2}+
    \right.\\
    &
    \left.
    \frac{(1-\rho^{k})\rho(\rho^{k}-\rho^{2n-k})(1+\rho+\rho^2-\rho^{2n-2k+2}(1+\rho^{k}+\rho^{2k}))}{(1-\rho)^2(1+\rho)(1+\rho+\rho^2)}
    \right)h
    +4e^{2at_p}\frac{\sigma^2}{2a}\\
    &
    \left(
    \frac{\rho^{4}(1-\rho^{4k})(1-\rho^{2n-2k})^2}{(1-\rho^2)^2(1+\rho^2)}
    + \frac{\rho^{2k}-(2n-2k+1)\rho^{2n}(1-\rho^2)-\rho^{4n-2k+2}}{(1-\rho^2)^2}
    \right)h^2,
    \end{aligned}
    \\[1ex]
    &\begin{aligned}
    \label{CovYZ sde 1}
    \mbox{Cov}(Y^{sde}_1,Z^{sde}_1) 
    = & \mbox{Cov}(Y^{sde},Z^{sde}) + 2(x_0 -b)e^{at_p} 
    \frac{\sigma^2}{2a} f_1(n,k,\rho) h 
     + 2 e^{2at_p} \frac{\sigma^2}{2a} f_2(n,k,\rho) h^2, \\
    \end{aligned}
\end{align}
where 
\begin{align*}
    f_1(n,k,\rho)= &
    \frac{\rho^{2k}-\rho^{2n}}{1-\rho^2}\left(
    3k\rho + 4\rho^3 +\frac{4(\rho-\rho^{2k}+\rho^{2n})-3\rho^2-\rho^{2k-1}+4\rho^{2k+1}}{1-\rho^2} 
    \right) -\frac{(1-\rho^{2k})(\rho^{2k-1}-\rho^{2n+1})}{(1-\rho^2)^2}\\
    &-2(n-k)\rho^{2n-1}\left(
    \frac{1+\rho^2}{1-\rho^{2}}+2\rho^{2}
    \right) + k\left(
    \frac{\rho^{2k-1}-\rho^{2n+1}}{1-\rho^{2}}+\rho^{2k-1}-\rho^{2n-1}
    \right)\\
    f_2(n,k,\rho)= & \frac{\rho^{2k}-\rho^{2n}}{1-\rho^2}\left(
    \frac{2\rho}{1-\rho}+2\rho(1+\rho^2)-\rho^{2k-1}-\frac{2\rho(\rho^{2k}+\rho^{2n})}{1+\rho^2}
     \right)
    -(n-k)\left(
    \rho^{2n-1}+2\rho^{2n+1}\frac{2-\rho^2}{1-\rho^2}
    \right).
\end{align*}
For a perturbation at $ t_p \in (t_{n-1}, t_n]$ (only the last observation is affected), the moments are
\begin{align*}
    & \mathbb{E}(Y^{sde}_1) = \mathbb{E}(Y),\\
    & \mathbb{E}(Z^{sde}_1) = \mathbb{E}(Z) + (x_0-b)e^{at_p}\rho^{2n-1}h,\\
    & V(Y^{sde}_1)  =  V(Y), \\
    & \mbox{Cov}(Y^{sde}_1 ,Z^{sde}_1) = \mbox{Cov}(Y, Z)+ 2 \frac{\sigma^2}{2a} (x_0-b) e^{at_p} \rho^{2n-1}(n-\frac{1-\rho^{2n}}{1-\rho^2})h.
\end{align*}

\section*{Acknowledgments}
This work has received funding from the European Union’s Horizon 2020 research and innovation program under the
Marie Skłodowska-Curie grant agreement No 956107, "Economic Policy in Complex Environments (EPOC)".


\bibliographystyle{acm} 

\bibliography{refs} 

\begin{thebibliography}{10}

\bibitem{aron:1984}
{\sc Aron, J.~L., and Schwartz, I.~B.}
\newblock Seasonality and period-doubling bifurcations in an epidemic model.
\newblock {\em Journal of Theoretical Biology 110}, 4 (1984), 665--679.
\newblock doi:10.1016/S0022-5193(84)80150-2.

\bibitem{Bartlett:1949}
{\sc Bartlett, M.~S.}
\newblock Some evolutionary stochastic processes.
\newblock {\em Journal of the Royal Statistical Society. Series B (Methodological) 11}, 2 (1949), 211--229.

\bibitem{Cuenod:2011}
{\sc Cuenod, C.-A., Favetto, B., Genon-Catalot, V., Rozenholc, Y., and Samson, A.}
\newblock Parameter estimation and change-point detection from dynamic contrast enhanced mri data using stochastic differential equations.
\newblock {\em Mathematical Biosciences 233}, 1 (2011), 68--76.

\bibitem{denmarkpopulation}
{\sc Denmark, S.}
\newblock Development in the number of inhabitants since 2008, 2023.
\newblock \url{https://www.dst.dk/en/Statistik/emner/borgere/befolkning/befolkningstal} [Accessed: 2023 August 4].

\bibitem{ditlevsen2021}
{\sc Ditlevsen, S., and Lansky, P.}
\newblock Only through perturbation can relaxation times be estimated.
\newblock {\em Physical Review E 86}, 5, 1 (2012).

\bibitem{ditlevsen:2020}
{\sc Ditlevsen, S., Rubio, A.~C., and Lansky, P.}
\newblock Transient dynamics of {P}earson diffusions facilitates estimation of rate parameters.
\newblock {\em Communications in Nonlinear Science and Numerical Simulations 82\/} (2020), 105034.

\bibitem{Donnet:2010}
{\sc Donnet, S., Foulley, J.-L., and Samson, A.}
\newblock Bayesian analysis of growth curves using mixed models defined by stochastic differential equations.
\newblock {\em Biometrics 66}, 3 (2010), 733--741.

\bibitem{SSIrapport}
{\sc for matematisk modellering af~covid 19, E.}
\newblock Ekspertrapport af den 15.januar 2021. {P}rognoser for smittetal med fokus på udviklingen i cluster {B}.1.1.7.
\newblock Tech. rep., Statens Serum Institut, 2021.

\bibitem{fuchsbook}
{\sc Fuchs, C.}
\newblock {\em Inference for Diffusion Processes. With Applications in Life Sciences}.
\newblock Springer, 2013.

\bibitem{GuedjJ2007MLEi}
{\sc Guedj, J., Thiébaut, R., and Commenges, D.}
\newblock Maximum likelihood estimation in dynamical models of {HIV}.
\newblock {\em Biometrics 63}, 4 (2007), 1198--1206.

\bibitem{Heng:2020}
{\sc Heng, K., and Althaus, C.~L.}
\newblock The approximately universal shapes of epidemic curves in the susceptible-exposed-infectious-recovered ({SEIR}) model.
\newblock {\em Scientific Reports 10}, 1 (2020), 19365.
\newblock doi: 10.1038/s41598-020-76563-8.

\bibitem{Huang2020}
{\sc Huang, H., Handel, A., and Song, X.}
\newblock A bayesian approach to estimate parameters of ordinary differential equation.
\newblock {\em Computational Statistics 35\/} (09 2020).

\bibitem{Jain2017}
{\sc Jain, P., Kakade, S.~M., Kidambi, R., Netrapalli, P., and Sidford, A.}
\newblock Accelerating stochastic gradient descent for least squares regression.
\newblock {\em arXiv.Org\/} (2017).

\bibitem{kermack:1927}
{\sc Kermack, W.~O., and McKendrick, A.~G.}
\newblock A contribution to the mathematical theory of epidemics.
\newblock {\em Proceedings of the Royal Society of London. Series A, Containing Papers of a Mathematical and Physical Character 115}, 772 (1927), 700--721.

\bibitem{Kozyreff:2022}
{\sc Kozyreff, G.}
\newblock Asymptotic solutions of the {SIR} and {SEIR} models well above the epidemic threshold.
\newblock {\em IMA Journal of Applied Mathematics\/} (2022), CorpusID:251116284.

\bibitem{Leander:2014}
{\sc Leander, J., Lundh, T., and Jirstrand, M.}
\newblock Stochastic differential equations as a tool to regularize the parameter estimation problem for continuous time dynamical systems given discrete time measurements.
\newblock {\em Mathematical Biosciences 251\/} (2014), 54--62.
\newblock doi: 10.1016/j.mbs.2014.03.001.

\bibitem{li2022constrained}
{\sc Li, D., and Wang, Y.}
\newblock Constrained unscented kalman filter for parameter identification of structural systems.
\newblock {\em Structural Control and Health Monitoring 29}, 4 (2022), e2908.

\bibitem{Liu:2020}
{\sc Liu, X., Xiao, T., Si, S., Cao, Q., and Hsieh, S. K. C.-J.}
\newblock How does noise help robustness? {E}xplanation and exploration under the neural {SDE} framework.
\newblock {\em 2020 IEEE/CVF Conference on Computer Vision and Pattern Recognition (CVPR)\/} (2020), 279--287.
\newblock doi: 10.1109/CVPR42600.2020.00036.

\bibitem{Nocedal:2006}
{\sc Nocedal, J., and Wright, S.~J.}
\newblock {\em Numerical Optimization}, 2~ed.
\newblock Springer, New York, NY, USA, 2006.
\newblock doi: 10.1007/978-0-387-40065-5.

\bibitem{OuLu2023Eonm}
{\sc Ou, L., Hunter, M.~D., Lu, Z., Stifter, C.~A., and Chow, S.-M.}
\newblock Estimation of nonlinear mixed-effects continuous-time models using the continuous-discrete extended {K}alman filter.
\newblock {\em British Journal of Mathematical \& Statistical Psychology 76}, 3 (2023), 462--490.

\bibitem{SPieschner}
{\sc Pieschner, S., Hasenauer, J., and Fuchs, C.}
\newblock Identifiability analysis for models of the translation kinetics after mrna transfection.
\newblock {\em Journal of Mathematical Biology 84\/} (05 2022).

\bibitem{pilipovic:2021}
{\sc Pilipovic, P., Samson, A., and Ditlevsen, S.}
\newblock Parameter estimation in nonlinear multivariate stochastic differential equations based on splitting schemes.
\newblock {\em The Annals of Statistics 52}, 2 (2024), 842--867.

\bibitem{Piovella:2020}
{\sc Piovella, N.}
\newblock Analytical solution of {SEIR} model describing the free spread of the {COVID}-19 pandemic.
\newblock {\em Chas, Solitons and Fractals 140\/} (2020), 110243.
\newblock doi: 10.1016/j.chaos.2020.110243.

\bibitem{R}
{\sc {R Core Team}}.
\newblock {\em R: A Language and Environment for Statistical Computing}.
\newblock R Foundation for Statistical Computing, Vienna, Austria, 2025.

\bibitem{SDEbook2019}
{\sc Särkkä, S., and Solin, A.}
\newblock {\em Applied Stochastic Differential Equations}.
\newblock Cambridge University Press, 2019.

\bibitem{TO2020565}
{\sc To, K. K.-W., Tsang, O. T.-Y., Leung, W.-S., Tam, A.~R., Wu, T.-C., Lung, D.~C., Yip, C. C.-Y., Cai, J.-P., Chan, J. M.-C., Chik, T. S.-H., Lau, D. P.-L., Choi, C. Y.-C., Chen, L.-L., Chan, W.-M., Chan, K.-H., Ip, J.~D., Ng, A. C.-K., Poon, R. W.-S., Luo, C.-T., Cheng, V. C.-C., Chan, J. F.-W., Hung, I. F.-N., Chen, Z., Chen, H., and Yuen, K.-Y.}
\newblock Temporal profiles of viral load in posterior oropharyngeal saliva samples and serum antibody responses during infection by sars-cov-2: an observational cohort study.
\newblock {\em The Lancet Infectious Diseases 20}, 5 (2020), 565--574.

\bibitem{Transtrum2012}
{\sc Transtrum, M.~K., and Sethna, J.~P.}
\newblock Improvements to the {L}evenberg-{M}arquardt algorithm for nonlinear least-squares minimization.
\newblock {\em arXiv.Org\/} (2012), 10.48550/arxiv.1201.5885.

\bibitem{10.5555/3157096.3157142}
{\sc Zhu, R.}
\newblock Gradient-based sampling: an adaptive importance sampling for least-squares.
\newblock In {\em Proceedings of the 30th International Conference on Neural Information Processing Systems\/} (2016), NIPS'16, Curran Associates Inc., p.~406–414.

\end{thebibliography}

\newpage

\setcounter{page}{1}
\section*{Supplementary material}

\renewcommand\thefigure{S\arabic{figure}}    
\setcounter{figure}{0}

Here are further figures exploring other parameter values in the simulation studies and supplementing figures in the main manuscript.

\subsection*{Example 1}

\begin{figure}[ht]
    \centering
    \includegraphics[width = 0.8\textwidth]{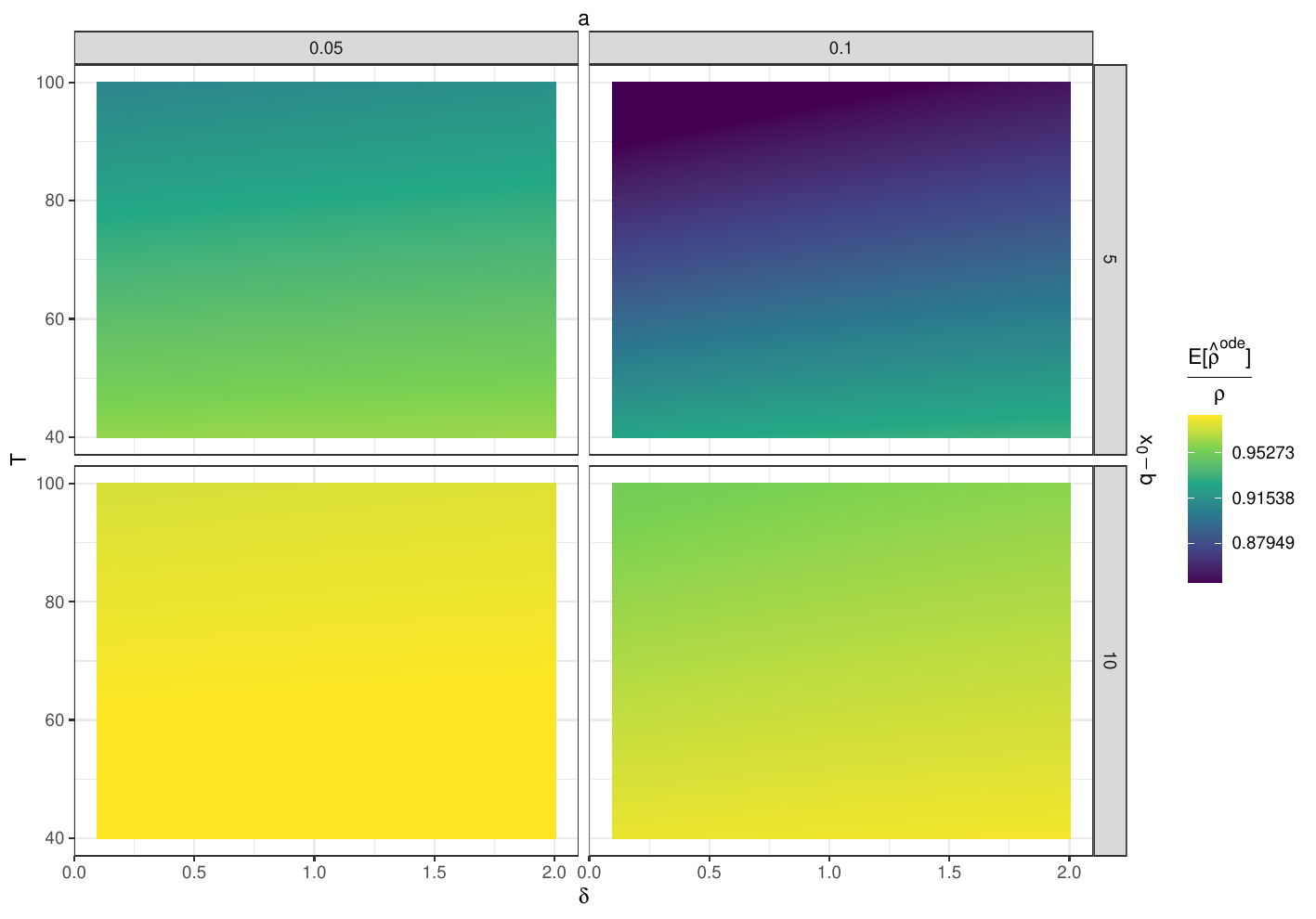} 
    \caption{The approximation of $\mathbb{E}(\hat{\rho}^{ode\cdot sde})/\rho$ in eq. \eqref{E hat rho 0} as a function of the observation interval $T$, time step $\Delta$, rate parameter $a$ ($a = 0.05$ in left column, $0.1$ in right column) and the initial value  $x_0 - b$ ($x_0= 5$ in upper row, $10$ in lower row). Other parameters are $b = 0$, $\sigma_0 = 0.1$. 
    }
    \label{fig: approximate bias MLE ODE}
\end{figure}

\begin{figure}[!]
    \centering
    \includegraphics[width = 0.8\textwidth]{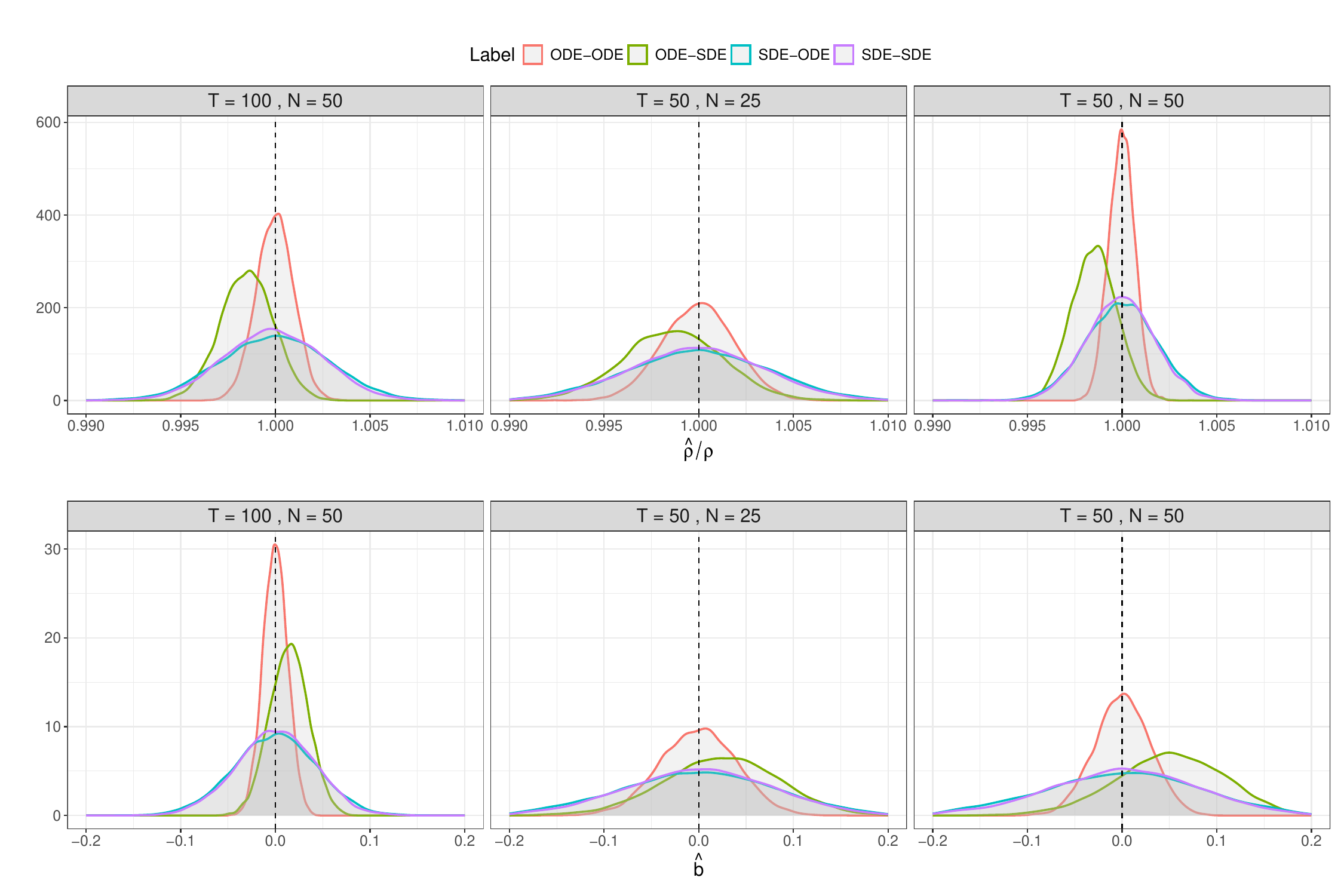} 
    \caption{{\bf Linear model without perturbation with $a = 0.05$.} Density plots of parameter estimates based on 10,000 simulated data sets of models \eqref{example 1 ODE} and \eqref{example 1 SDE}. The labels indicate (data-generating model) - (model used for estimation). 1st column: $T=100$, $n=50$; 2nd column: $T=50$, $n=50$; 3rd column: $T=50$, $n=25$. Upper row: $\hat \rho/\rho$. Lower row: $\hat b$. The black dashed lines indicate the true values. Parameters are $a = 0.05, b=0, \sigma_0 = 0.05$ and $\sigma = \sigma_0 \sqrt{2a}$.}
   \label{fig: basic model simulation 1}
\end{figure}

\begin{figure}[ht]
    \centering
    \includegraphics[width = 0.8\textwidth]{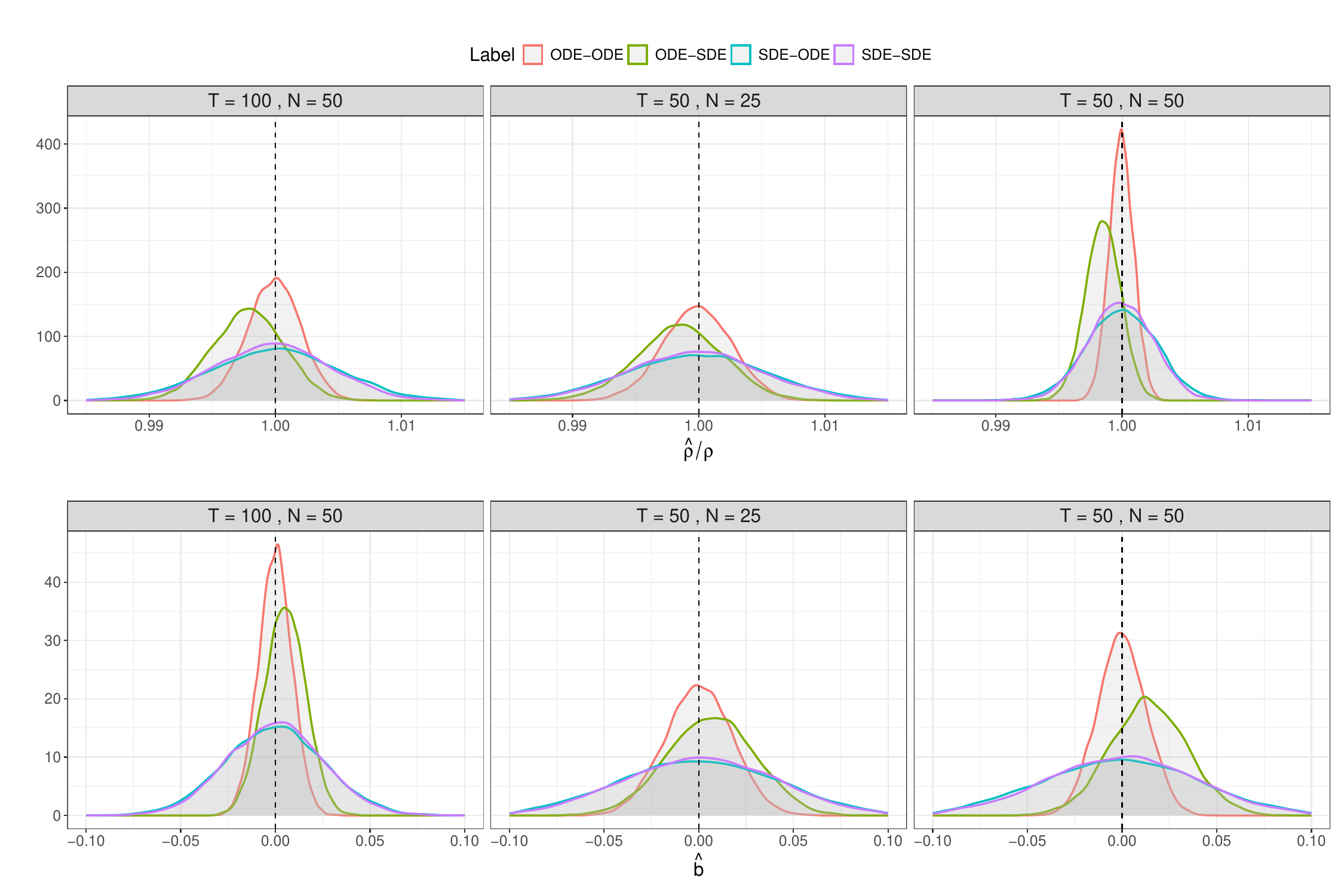} 
    \caption{{\bf Linear model with no perturbation with $a = 0.1$.} Density plots of parameter estimates based on 10,000 simulated data sets of models \eqref{example 1 ODE} and \eqref{example 1 SDE}. The labels indicate (data-generating model) - (model used for estimation). 1st column: $T=100$, $n=50$; 2nd column: $T=50$, $n=50$; 3rd column: $T=50$, $n=25$. Upper row: $\hat \rho/\rho$. Lower row: $\hat b$. The black dashed lines indicate the true values $a = 0.1, b=0$. Other parameters are $\sigma_0 = 0.05$ and $\sigma = \sigma_0 \sqrt{2a}$.}
   \label{fig: basic model simulation 2}
\end{figure}

\begin{figure}[ht]
    \centering
    \includegraphics[width = 0.8\textwidth]{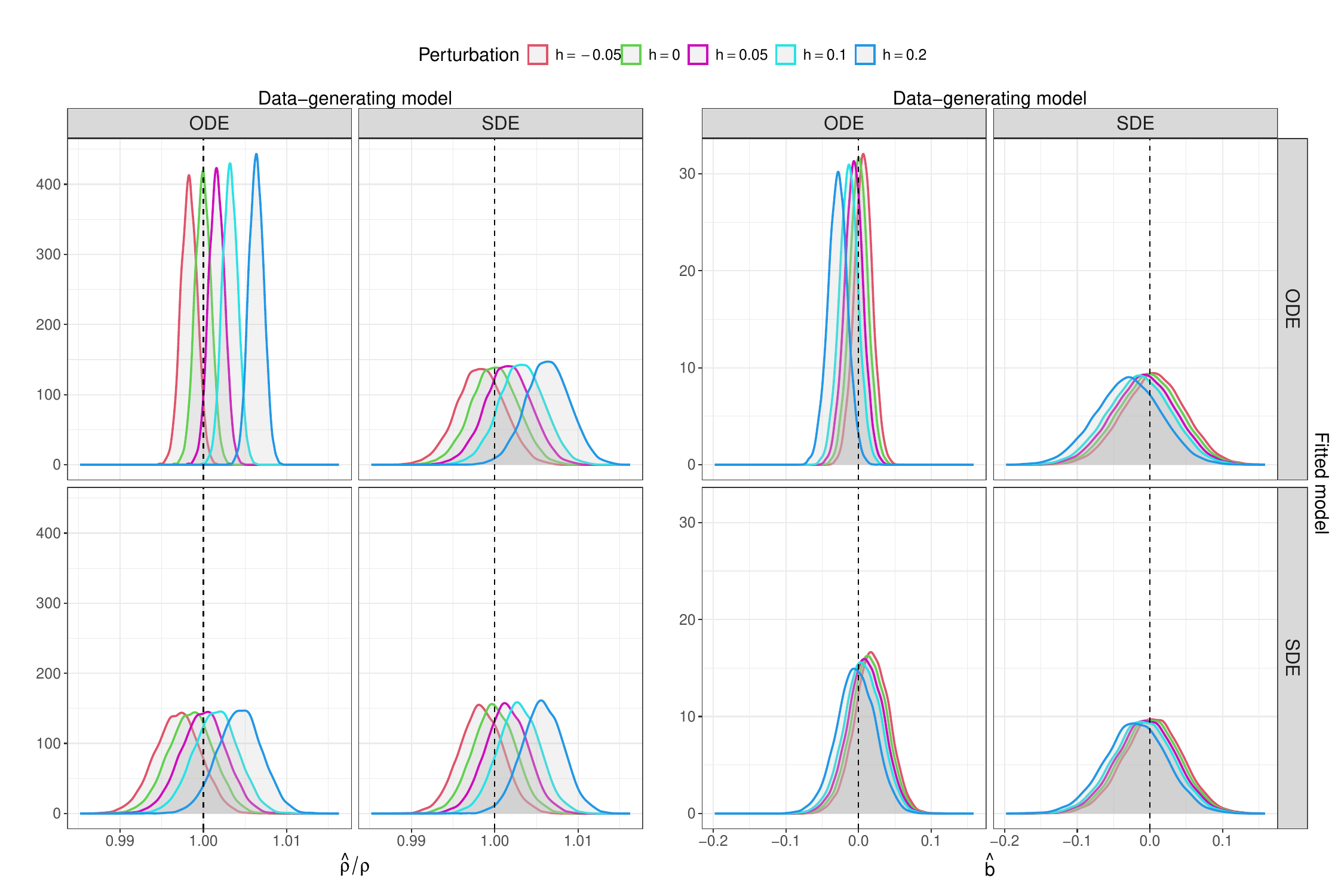} 
    \caption{{\bf Linear model with instantaneous perturbation.} Density plots of parameter estimates based on 10,000 simulated data sets of models \eqref{example 1 ODE} and \eqref{example 1 SDE} with instantaneous perturbation \eqref{model instantaneous} with $T=100$ and different values of $h$. Using data from \texttt{Data-generating model} (columns), we fit it to the \texttt{Fitted model} (rows). The black dashed lines indicate the true values. Parameters are $a = 0.05, b=0$, $\sigma_0 = 0.05$ and $\sigma = \sigma_0 \sqrt{2a}$. }
    \label{fig: instantaneous perturbation sim2}
\end{figure}

\begin{figure}[ht]
    \centering
    \includegraphics[width = 0.8\textwidth]{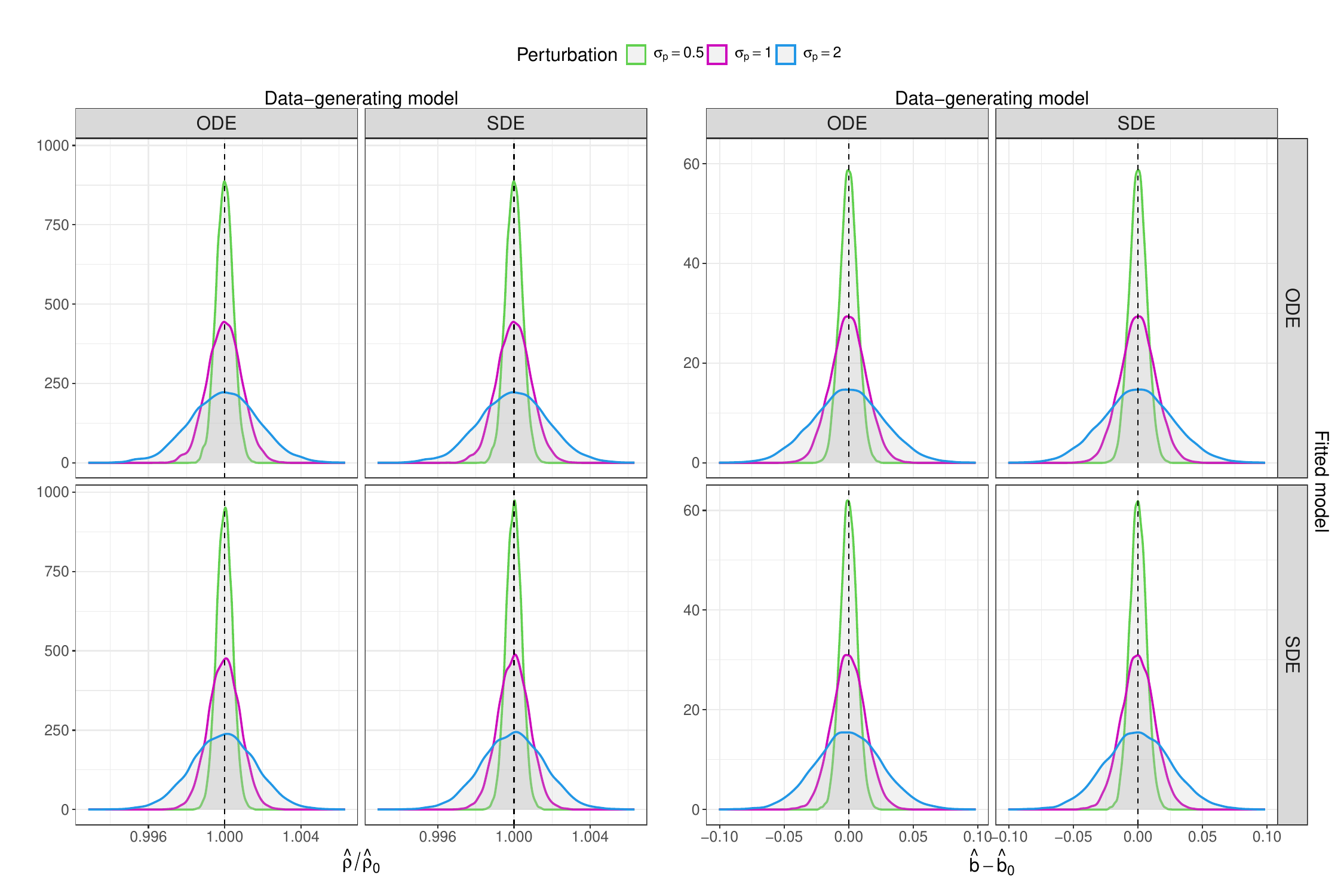} 
    \caption{{\bf Linear model with randomly varying long-term mean.} Density plots of parameter estimates based on 10,000 simulated data sets of models \eqref{example 1 ODE} and \eqref{example 1 SDE} with randomly varying long-term mean \eqref{The oscillation of the long-term mean} with $T=100$, $\Delta = 2$ and different values of $\sigma_b$. Using data from \texttt{Data-generating model} (columns), we fit it to the \texttt{Fitted model} (rows). The black dashed lines indicate the true values. Parameters are $a = 0.05$, b=0, $\sigma_0 = 0.05$ and $\sigma = \sigma_0 \sqrt{2a}$.}
    \label{fig: oscillation simulation 1}
\end{figure}

\begin{figure}[ht]
    \centering
    \includegraphics[width = 0.8\textwidth]{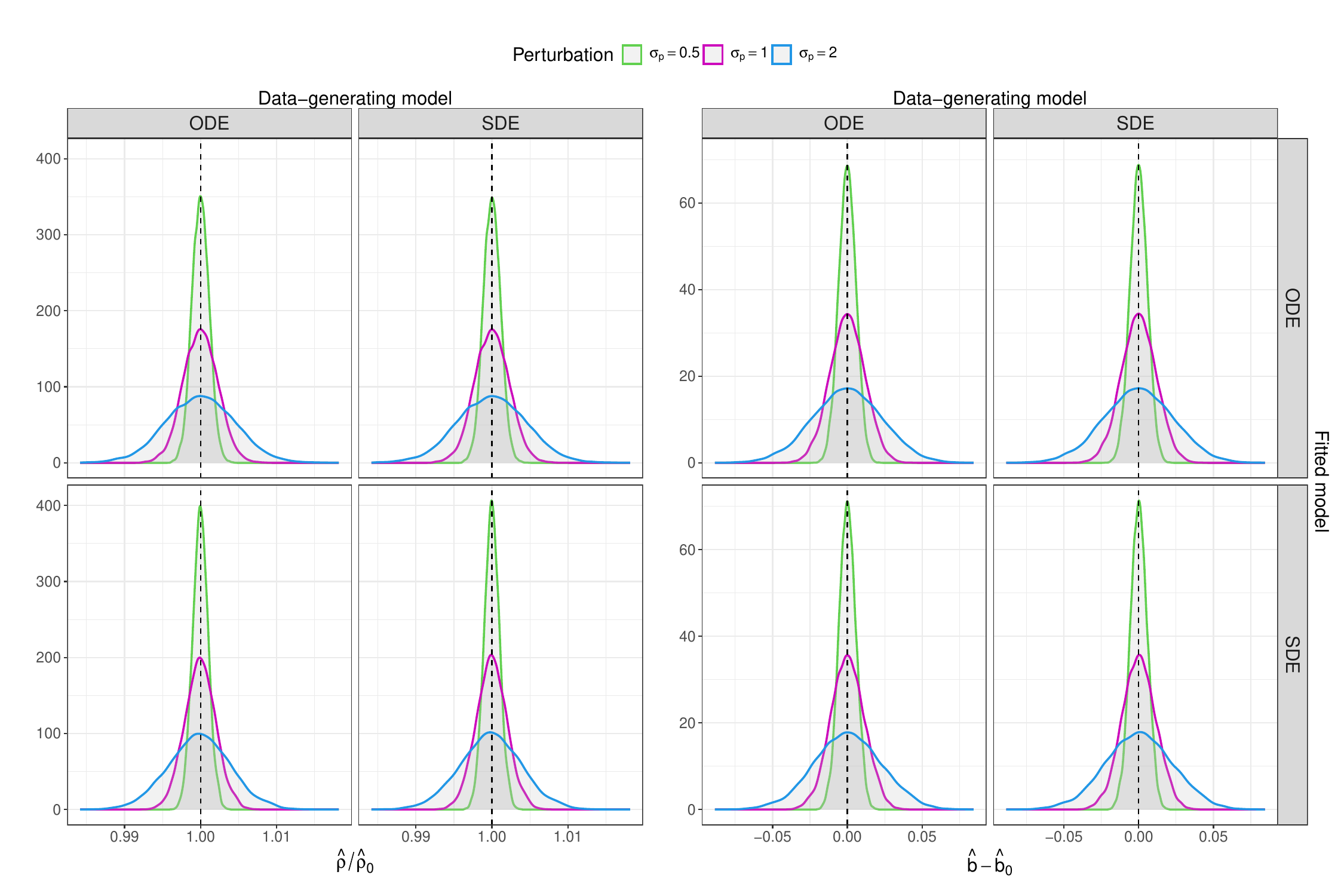} 
    \caption{{\bf Linear model with randomly varying long-term mean.} Density plots of parameter estimates based on 10,000 simulated data sets of models \eqref{example 1 ODE} and \eqref{example 1 SDE} with randomly varying long-term mean \eqref{The oscillation of the long-term mean} with $T=100$, $\Delta = 2$ and different values of $\sigma_b$. Using data from \texttt{Data-generating model} (columns), we fit it to the \texttt{Fitted model} (rows). The black dashed lines indicate the true values. Parameters are $a = 0.1, b=0$, $\sigma_0 = 0.05$ and $\sigma = \sigma_0 \sqrt{2a}$.}
    \label{fig: oscillation simulation 2}
\end{figure}

\begin{figure}[ht]
    \centering
    \includegraphics[width = 0.8\textwidth]{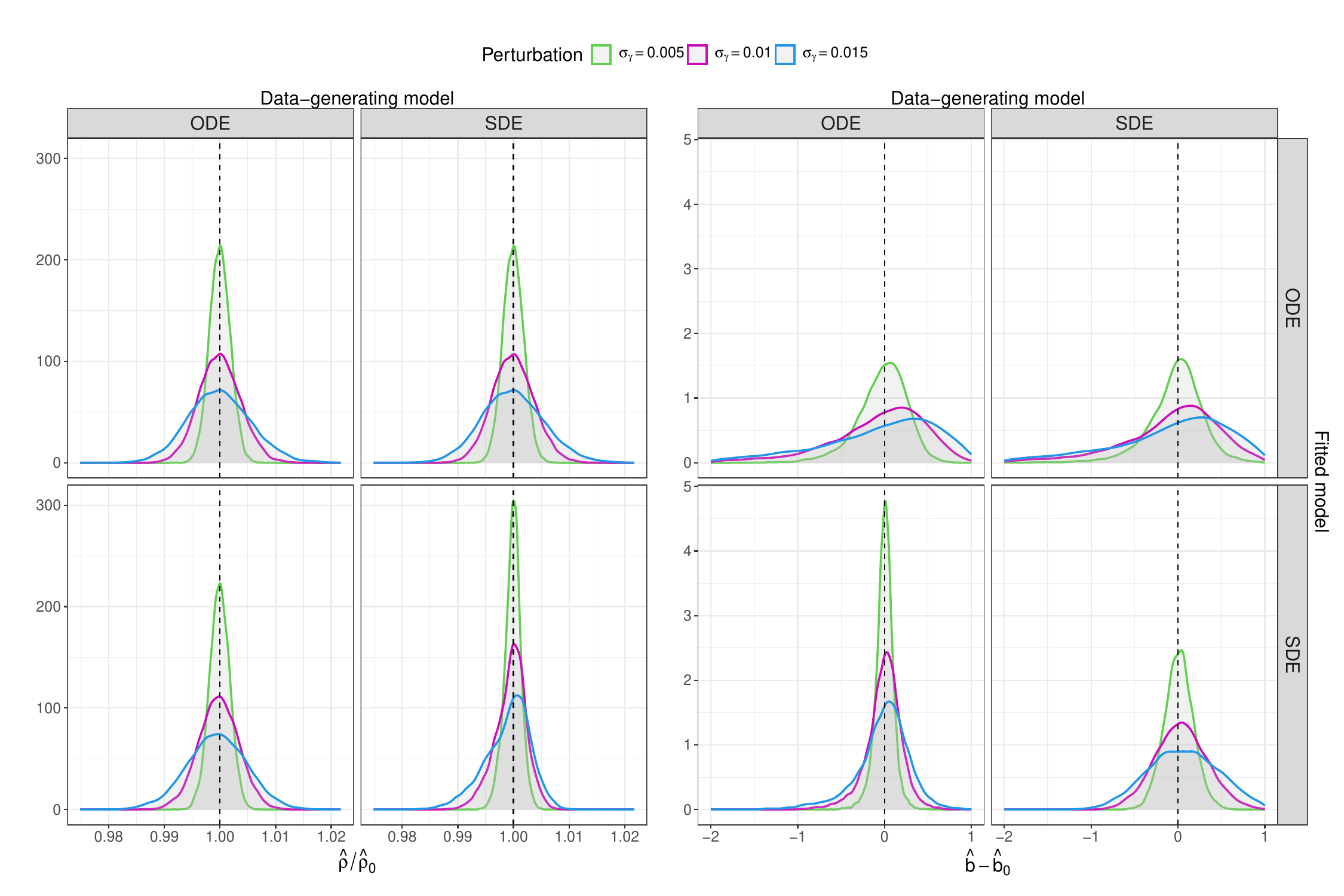} 
    \caption{{\bf Linear model with quadratic misspecification term.} Parameter estimates from 10,000 simulated data sets. Models with squared error term $\gamma_t (x_t-b)^2$ were simulated from models \eqref{example 1 ODE} (ODE) or \eqref{example 1 SDE} (SDE). Data from the \texttt{Data-generating model} (columns), were fitted to the \texttt{Fitted model} (rows). The black dashed lines indicate the true values. Parameters are $a = 0.01, x_0 = 5$, $\sigma_0 = 0.05$ and $\sigma = \sigma_0 \sqrt{2a}$.
    }
    \label{fig: Error term in differential equations sim2}
\end{figure}

\begin{figure}[!ht]
    \centering
   \includegraphics[width = 0.8\textwidth]{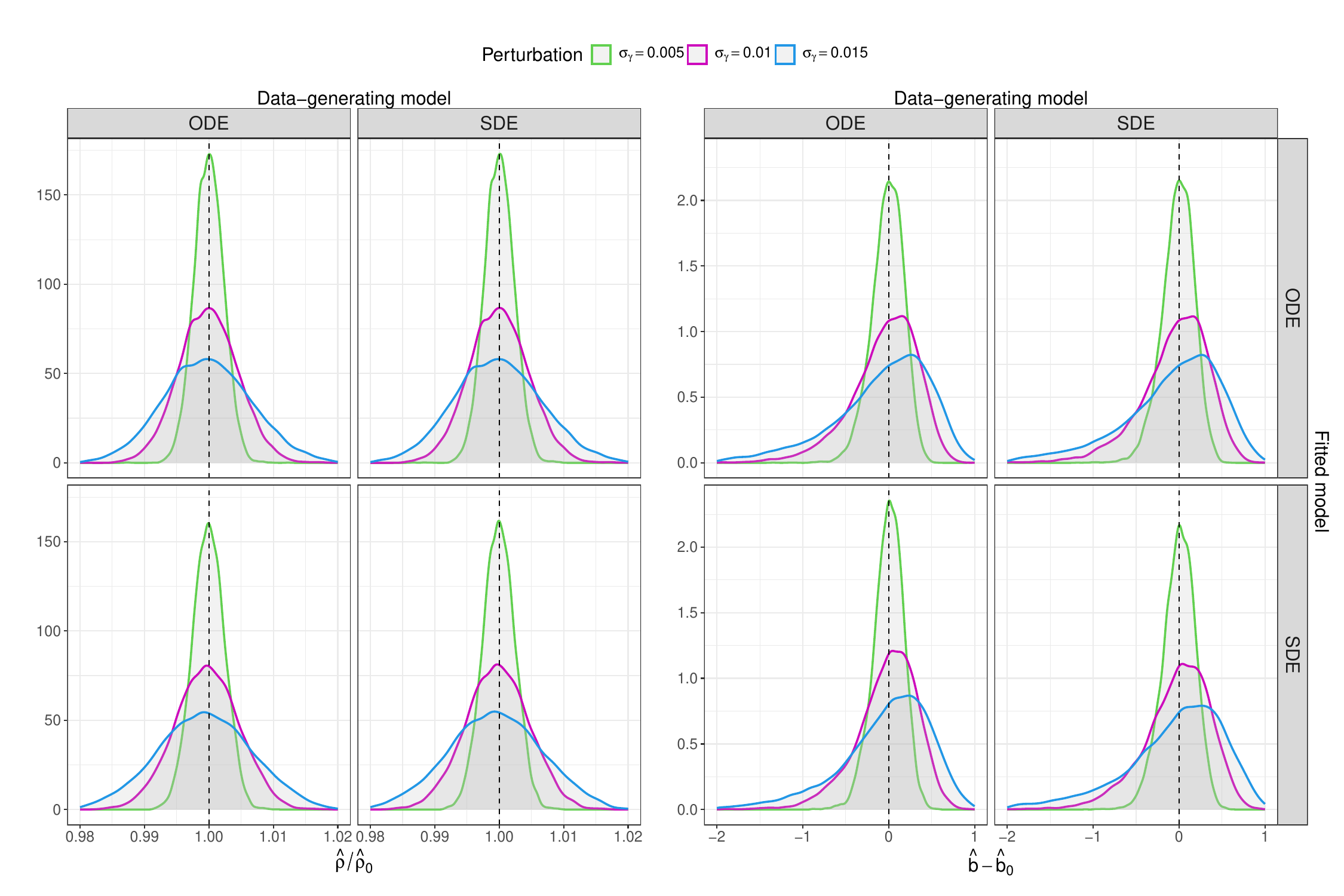} 
    \caption{{\bf Linear model with quadratic misspecification term.} Parameter estimates from 10,000 simulated data sets. Models with squared error term $\gamma_t (x_t-b)^2$ were simulated from models \eqref{example 1 ODE} (ODE) or \eqref{example 1 SDE} (SDE). Data from the \texttt{Data-generating model} (columns), were fitted to the \texttt{Fitted model} (rows). The black dashed indicate the true values. Parameters are $a = 0.02, x_0 = 10$, $\sigma_0 = 0.05$ and $\sigma = \sigma_0 \sqrt{2a}$.}
    \label{fig: Error term in differential equations sim3}
\end{figure}

\end{document}